\documentclass[aps,showpacs,eqsecnum,twocolumn,superscriptaddress,floatfix]{revtex4}

\usepackage{amsmath}
\usepackage{latexsym}
\usepackage{times}
\usepackage{amssymb}

\usepackage{ifpdf}
\usepackage{graphicx}
\newcommand{\imgname}[1]{#1.eps}
\newcommand{\COLORimgname}[1]{COLOR#1.eps}

\ifpdf
\fi

\begin{document}

\title[Accurate simulations of the dynamical bar-mode instability in
  full General Relativity]{Accurate simulations of the dynamical
  bar-mode instability in full General Relativity}

\author{Luca Baiotti}
\affiliation{Max-Planck-Institut f\"ur Gravitationsphysik,
Albert-Einstein-Institut, Golm, Germany}
\author{Roberto \surname{De Pietri}}
\author{Gian Mario \surname{Manca}}
\affiliation{Dipartimento di Fisica, 
  Universit\`a di Parma and INFN, Parma, Italy}
\author{Luciano Rezzolla}
\affiliation{Max-Planck-Institut f\"ur Gravitationsphysik,
Albert-Einstein-Institut, Golm, Germany}
\affiliation{SISSA, International School for Advanced Studies
and INFN, Trieste,  Italy}
\affiliation{Department of Physics,  Louisiana State University, 
Baton Rouge, USA}

\date{\today}
\begin{abstract}
  We present accurate simulations of the dynamical bar-mode instability in
  full General  Relativity focussing on  two aspects which have  not been
  investigated in detail  in the past. Namely, on  the persistence of the
  bar deformation once the instability  has reached its saturation and on
  the  precise  determination of  the  threshold  for  the onset  of  the
  instability in terms of  the parameter $\beta={T}/{|W|}$.  We find that
  generic nonlinear  mode-coupling effects appear  during the development
  of the instability and these  can severely limit the persistence of the
  bar deformation  and eventually suppress the  instability. In addition,
  we observe the dynamics of the instability to be strongly influenced by
  the  value  $\beta$ and  on  its  separation  from the  critical  value
  $\beta_c$ marking the onset of  the instability.  We discuss the impact
  these results  have on the  detection of gravitational waves  from this
  process and  provide evidence that the  classical perturbative analysis
  of the  bar-mode instability for Newtonian  and incompressible Maclaurin
  spheroids remains qualitatively valid and accurate also in full General
  Relativity.
\end{abstract}

\maketitle

\section{Introduction}
\label{sec:intro}

It is well known that rotating neutron stars are subject to
non-axisymmetric instabilities for non-radial axial modes with azimuthal
dependence $e^{i m \phi}$ (with $m=1,2,\ldots$) when the instability
parameter $\beta\equiv T/|W|$ (\textit{i.e.} the ratio between the
rotational kinetic energy $T$ and the gravitational binding energy $W$)
exceeds a critical value $\beta_{c}$.

An exact and perturbative treatment of these instabilities exists only
for incompressible self-gravitating fluids in Newtonian gravity (see
refs.~\cite{Chandrasekhar69,ShapiroTeukolsky83}) and this predicts that a
dynamical instability should arise for the ``bar-mode'' (\textit{i.e.}
the one with $m$=2) when $\beta \geq \beta_{c} = 0.2738$. On the other
hand, the accurate study of the dynamical bar-mode instability, of its
nonlinear evolution and of the determination of the threshold for the
instability, demand the use of numerical simulations with the solution in
three spatial dimensions (3D) of the fully nonlinear hydrodynamical
equations coupled to the Einstein field equations.

Despite these requirements, much of the literature on this process has so
far been limited to a Newtonian or post-Newtonian (PN) description. While
this represents an approximation, these studies have provided important
information on several aspects of the instability that could not have
been investigated with perturbative techniques. In particular, these
numerical studies have shown that $\beta_{c}$ depends very weakly on the
stiffness of the equation of state (EOS) and that, once a bar has
developed, the formation of spiral arms is important for the
redistribution of the angular momentum (see refs.~\cite{Tohline85,
Tohline86, Brown:2000, HouserCentrellaSmith94, HouserCentrella96,
SmithHouserCantrella96, NewCentrellaTohline00, NewShapiro2001,
Liu2002}). More recently, instead, it was shown that the threshold for
the onset of the dynamical instability can be smaller for stars with a
high degree of differential rotation and a weak dependence on the EOS was
confirmed in refs.~\cite{PicketDurisenDavis96, TohlineHachisu90,
KarinoEriguchi03, ShibataKarinoEriguchi2002,
ShibataKarinoEriguchi2003}. Finally, these Newtonian analyses have also
provided the first evidence that an $m$=1-mode dynamical instability may play
an important role for smaller values of the critical parameter. This is
also referred to as the ``low-$\beta$''
instability~\cite{Saijo:2003,Ott05} and it will not be considered here.

Only very recently it has become possible to perform simulations of the
dynamical bar instability for old neutron stars in full General
Relativity~\cite{Shibata00b}. These studies have shown that within a
fully general-relativistic framework the critical value for the onset of
the instability is smaller than the Newtonian one (\textit{i.e.}
$\beta_{c}\simeq 0.24-0.25$) and this behaviour was confirmed by PN
calculations~\cite{Saijo2000,Saijo:2001} which also suggested that
$\beta_{c}$ varies with the compactness $M/R$ of the star.

The bar-mode instability may take place in young neutron stars, either
as the result of the accretion-induced collapse of a white
dwarf~\cite{Dessart2006} or in the collapse of a massive stellar
core. Indeed, recent simulations investigating axisymmetric
stellar-core collapse in full General
Relativity~\cite{ShibataSekiguchi2004} have pointed out that for
sufficiently differentially rotating progenitors, it is, at least in
principle, possible to obtain toroidal protoneutron-star cores with
masses between 1.2 and 2.1 $M_\odot$ which are unstable against bar-mode
deformations (it should be mentioned that it is still unclear how likely
these high rotation-rates and strongly differential-rotation profiles
actually are in nature). Besides pointing out this alternative
interesting scenario, the work in ref.~\cite{ShibataSekiguchi2004} has
also suggested that more realistic EOS and neutrino cooling could
enhance the process; this scenario has also been considered in the
PN approximation in ref.~\cite{Saijo2005}. An important
common feature in all these investigations is that the development of
the bar-mode instability was obtained introducing very strong
\textit{ad hoc} $m$=2-mode perturbations.

The recent general-relativistic studies, together with their Newtonian
and PN counterparts, have been very helpful in
highlighting the main features of the instability. However, several
fundamental questions remain unanswered. Most notably: \textit{i)}
What is the role of the initial perturbation?  \textit{ii)} What is the
effect of the symmetry conditions often used in numerical
calculations? {\it iii)} How are the dynamics influenced by the value
of the parameter $\beta$, especially when this is largely
overcritical?  Finally and most importantly: \textit{iv)} How long
does a bar survive, once fully developed?

Clearly, the last question has important implications for the possible
observational relevance of the gravitational-wave signal emitted through
the bar-mode instability as the signal-to-noise ratio (SNR) can increase
considerably in the case of a long-lived bar since the SNR grows as the
square root of the number of the effective cycles of the signal available
for detection. Earlier work on this subject basically suggested that once
a bar was formed it would tend to be persistent on the radiation-reaction
timescale. Indications and evidences in this direction were presented in
a Newtonian framework in ref.~\cite{Brown:2000} as well as in a PN one in
ref.~\cite{Saijo:2001}. In contrast, fully general-relativistic results,
either using suitable symmetry boundary conditions (\textit{i.e.} the so
called $\pi$-symmetry boundary conditions)~\cite{Shibata00b} or
not~\cite{ShibataSekiguchi2004}, show a non-persistent bar.

In this work we try to find answers to these important open questions
by exploring, in a systematic way, the bar-mode instability for a large
number of initial stellar models. In doing this, we intend to go beyond
the standard phenomenological discussion of the nonlinear dynamics of
the instability often encountered in the literature.

The main results of our analysis can be summarised as follows:
\textit{i)} The initial perturbation (either in the form of an $m$=1 mode
or of an $m$=2 mode) can play a role in determining the duration of the
bar-mode deformation, but not in determining the growth time of the
instability; the only exception to this is represented by models near 
the threshold; \textit{ii)} For moderately overcritical models
(\textit{i.e.} with $\beta \approxeq \beta_c$), the use of a
$\pi$-symmetry can radically change the dynamics and extend considerably
the persistence of the bar; this ceases to be true for largely
overcritical models (\textit{i.e.} with $\beta \gg \beta_c$), for which
even the artificial symmetries are not sufficient to provide a long-lived
bar; \textit{iii)} The persistence of the bar is strongly dependent on
the degree of overcriticality and is generically of the order of the
dynamical timescale; \textit{iv)} Generic nonlinear mode-coupling effects
(especially between the $m$=1 and the $m$=2 modes) appear during the
development of the instability and these can severely limit the
persistence of the bar deformation and eventually suppress the
{bar deformation}; \textit{v)} The dynamics of
largely overcritical models are fully determined by the excess of
rotational energy and the bar deformation is very rapidly suppressed
through the conversion of kinetic energy into internal one. In addition,
we have also assessed the accuracy of the classical Newtonian stability
analysis of Maclaurin spheroids for incompressible self-gravitating
fluids~\cite{Chandrasekhar69}. Overall, and despite having applied it to
differentially rotating and relativistic models, we have found it to be
surprisingly accurate in determining both the threshold for the
instability and the complex eigenfrequencies for the unstable models.

The paper is organized as follows. In Sec.~II we give details on the
evolution methods used, while in Sec.~III we discuss the initial
models and their properties. In Sec.~IV we introduce the methodology
used to analyse the numerical results of the simulations, which are
then discussed in Sec.~V in terms of the general dynamics of the
instability and of the general properties. In Sec.~VI, we present the 
features of the instability that are specific to
different treatments of the initial conditions, while in
Sec.~VII we illustrate two different methods for the determination of
$\beta_c$. Finally, in Sec.~VIII we discuss the impact of our results
on the emission of gravitational waves from the unstable models and
present in Sec.~IX our conclusions and the prospects of future
research.

We have used a space like signature $(-,+,+,+)$, with Greek indices
running from 0 to 3, Latin indices from 1 to 3 and the standard
convention for the summation over repeated indices. Furthermore, we indicate as ($x$,$y$,$z$) the
Cartesian coordinates and we define $r=\sqrt{x^2+y^2+z^2}$, $\varpi=\sqrt{x^2+y^2}$,
$\theta=\arctan(\varpi/z)$, $\phi=\arctan(y/x)$ for the axial and
spherical coordinates. Unless explicitly stated, all the
quantities are expressed in the system of adimensional units in which
$c=G=M_\odot=1$.

\section{Evolution of Fields and Matter}
\label{sec:evolution}

The code and the evolution method are the same as the ones used in
Baiotti et al.~\cite{Baiotti03a,Baiotti04a} and therein described. For
convenience we report here the main properties and characteristics of the
employed simulation method. We have used the general-relativistic
hydrodynamics code {\tt Whisky}, in which the hydrodynamics equations are
written as finite differences on a Cartesian grid and solved using
high-resolution shock-capturing (HRSC) schemes (a first description of
the code was given in~\cite{Baiotti04a}).

\subsection{Evolution of Einstein equations}    
\label{sec:Evol_EINSTEIN_EQ}

The original ADM formulation casts the Einstein equations into a
first-order (in time) quasi-linear~\cite{Richtmyer67} system of
equations. The dependent variables are the three-metric $\gamma_{ij}$ and
the extrinsic curvature $K_{ij}$, with first-order evolution equations
given by
\begin{eqnarray}
\partial_t \gamma_{ij} &=& - 2 \alpha K_{ij}+\nabla_i
        \beta_j + \nabla_j \beta_i, 
\label{dtgij} \\
        \partial_t K_{ij} &=& -\nabla_i \nabla_j \alpha + \alpha \Biggl[
        R_{ij}+K\ K_{ij} -2 K_{im} K^m_j  \nonumber \\
        &\ & - 8 \pi \left( S_{ij} - \frac{1}{2}\gamma_{ij}S \right)
        - 4 \pi {\rho}_{_{\rm ADM}} \gamma_{ij}
        \Biggr] \nonumber \\ 
        &\ & + \beta^m \nabla_m K_{ij}+K_{im} 
        \nabla_j \beta^m+K_{mj} \nabla_i \beta^m .
        \nonumber \\
\label{dtkij}
\end{eqnarray}
Here, $\alpha$ is the lapse function, $\beta_i$ is the shift vector,
$\nabla_i$ denotes the covariant derivative with respect to the
three-metric $\gamma_{ij}$, $R_{ij}$ is the Ricci curvature of the
three-metric, $K\equiv\gamma^{ij}K_{ij}$ is the trace of the extrinsic
curvature, $S_{ij}$ is the projection of the stress-energy tensor onto
the space-like hypersurfaces and $S \equiv \gamma^{ij} S_{ij}$ (for a
more detailed discussion, see~\cite{York79}). In addition to the
evolution equations, the Einstein equations also provide four constraint
equations to be satisfied on each space-like hypersurface. These are the
Hamiltonian constraint equation
\begin{equation}
\label{ham_constr}
{}^{(3)}R + K^2 - K_{ij} K^{ij} - 16 \pi
        {\rho}_{_{\rm ADM}} = 0 \ ,
\end{equation}
and the momentum constraint equations
\begin{equation}
\label{mom_constr}
\nabla_j K^{ij} - \gamma^{ij} \nabla_j K - 8 \pi j^i = 0 \ .
\end{equation}
In equations (\ref{dtgij})--(\ref{mom_constr}), $ {\rho}_{_{\rm ADM}}$
and $j^i$ are the energy density and the momentum density as measured by
an observer moving orthogonally to the space-like hypersurfaces.

In particular, we use a conformal traceless reformulation of the above
system of evolution equations, as first suggested by Nakamura, Oohara and
Kojima~\cite{Nakamura:87} (NOK formulation), in which the evolved
variables are the conformal factor ($\phi$), the trace of the extrinsic
curvature ($K$), the conformal 3-metric $(\tilde{\gamma}_{ij})$, the
conformal traceless extrinsic curvature $(\tilde{A}_{ij})$ and the {\it
conformal connection functions } $(\tilde{\Gamma}^i)$, defined as
\begin{eqnarray}
\phi &=& \frac{1}{4} \log(\sqrt[3]{\gamma})  \quad ,\\
K    &=& \gamma^{ij}K_{ij}  \quad ,\\
\tilde{\gamma}_{ij} &=&e^{-4\phi}\gamma_{ij}\quad , \\
\tilde{A}_{ij}      &=&e^{-4\phi} ( K_{ij}-\gamma_{ij}K) \quad ,\\
\tilde{\Gamma}^i    &=&\tilde{\gamma}^{ij}_{,j} \quad .
\end{eqnarray}

The code used for evolving these quantities is the one developed within
the \texttt{Cactus} computational toolkit~\cite{Cactusweb} and is
designed to handle arbitrary shift and lapse conditions.  In particular,
we have used hyperbolic $K$-driver slicing conditions of the form
\begin{equation}
\partial_t \alpha = - f(\alpha) \;
\alpha^2 (K-K_0),
\label{eq:BMslicing}
\end{equation}
with $f(\alpha)>0$ and $K_0 \equiv K(t=0)$. This is a generalization of
many well known slicing conditions.  For example, setting $f=1$ we
recover the ``harmonic'' slicing condition, while, by setting
\mbox{$f=q/\alpha$}, with $q$ an integer, we recover the generalized
``$1+$log'' slicing condition~\cite{Bona94b}.  In particular, all the
simulations discussed in this paper are done using condition
(\ref{eq:BMslicing}) with $f=2/\alpha$. This choice has been made mostly
because of its computational efficiency, but we are aware that ``gauge
pathologies'' could develop with the ``$1+$log''
slicings~\cite{Alcubierre97a,Alcubierre97b}.

As for the spatial gauge, we use one of the ``Gamma-driver'' shift
conditions proposed in~\cite{Alcubierre01a}, that essentially acts so
as to drive the $\tilde{\Gamma}^{i}$ to be constant. In this respect,
the ``Gamma-driver'' shift conditions are similar to the
``Gamma-freezing'' condition $\partial_t \tilde\Gamma^k=0$, which, in
turn, is closely related to the well-known minimal distortion shift
condition~\cite{Smarr78b}.

In particular, all the results reported here have been obtained
using the hyperbolic Gamma-driver condition,
\begin{equation}
\partial^2_t \beta^i = F \, \partial_t \tilde\Gamma^i - \eta \,
\partial_t \beta^i,
\label{eq:hyperbolicGammadriver}
\end{equation}
where $F$ and $\eta$ are, in general, positive functions of space and
time. For the hyperbolic Gamma-driver conditions it is crucial to add
a dissipation term with coefficient $\eta$ to avoid strong
oscillations in the shift. Experience has shown that by tuning the
value of this dissipation coefficient it is possible to almost freeze
the evolution of the system at late times. We typically choose
$F=\frac{3}{4}\alpha$ and $\eta=2$ and do not vary them in time.

\subsection{Evolution of the hydrodynamics equations}    
\label{sec:hydroEV}

The stellar models are here treated in terms of a perfect fluid with
stress-energy tensor
\begin{align}
T^{\mu\nu} &= \rho h u^{\mu} u^{\nu} + p g^{\mu\nu} \, ,\\
h    &= 1 +\epsilon + \frac{p}{\rho} \quad ,
\end{align}
where $h$ is the specific enthalpy, $\epsilon$ the specific internal
energy and $\rho$ the rest-mass density, so that $e = \rho
(1+\epsilon)$ is the energy density in the rest-frame of the
fluid. The equations of relativistic hydrodynamics are then given by
the conservation laws for the energy, momentum and baryon number
\begin{equation}
\label{hydro eqs}
\begin{array}{lcl}
\nabla _{\mu} T^{\mu\nu}= 0 \ ,
&& 
\\[1mm]
\nabla _{\mu} (\rho u^{\mu}) = 0 \ ,
&& 
\end{array}
\end{equation}
once supplemented with an EOS of type $p =
p(\rho,\epsilon)$. While the code has been written to use any EOS, all
the simulations presented here have been performed using either an
isentropic ``polytropic'' EOS
\begin{equation}
\label{eq:EOSisentropic}
p = K \rho^{\Gamma}\ ,
\end{equation}
where $K$ is the polytropic constant and $\Gamma$ the adiabatic
exponent, or a non-isentropic ``ideal-fluid'' ($\Gamma$-law) EOS 
\begin{equation}
\label{eq:EOSideal}
p = (\Gamma-1) \rho\, \epsilon \ . 
\end{equation}
Note that, with the exception of the polytropic EOS
(\ref{eq:EOSisentropic}), the entropy is not constant and thus the
evolution equation for $\epsilon$ needs to be solved.

An important feature of the {\tt Whisky} code is the implementation of
a \textit{conservative formulation} of the hydrodynamics equations in
which the set of equations (\ref{hydro eqs}) is written in a
hyperbolic, first-order and flux-conservative form of the type
\begin{equation}
\label{eq:consform1}
\partial_t {\mathbf q} + 
        \partial_i {\mathbf f}^{(i)} ({\mathbf q}) = 
        {\mathbf s} ({\mathbf q})\ ,
\end{equation}
where ${\mathbf f}^{(i)} ({\mathbf q})$ and ${\mathbf s}({\mathbf q})$
are the flux-vectors and source terms, respectively (see
ref.~\cite{Font03} for an explicit form of the equations).  Note that
the right-hand side (the source terms) depends only on the metric, and
its first derivatives, and on the stress-energy tensor.  In order to
write system (\ref{hydro eqs}) in the form of system
(\ref{eq:consform1}), the \textit{primitive} hydrodynamical variables
(\textit{i.e.} the rest-mass density $\rho$ and the pressure $p$
(measured in the rest-frame of the fluid), the fluid three-velocity
$v^i$ (measured by a local zero--angular-momentum observer), the
specific internal energy $\epsilon$ and the Lorentz factor $W=\alpha u^0$) are
mapped to the so called \textit{conserved} variables \mbox{${\mathbf
    q} \equiv (D, S^i, \tau)$} via the relations
\begin{eqnarray}
\label{eq:prim2con}
   D &\equiv&  \sqrt{\gamma}W\rho\ , \nonumber\\
   S^i &\equiv& \sqrt{\gamma} \rho h W^2 v^i\ ,  \\
   \tau &\equiv& \sqrt{\gamma}\left( \rho h W^2 - p\right) - D\ . \nonumber
\end{eqnarray}
As previously noted, in the case of the polytropic EOS
\eqref{eq:EOSisentropic}, one of the evolution equations (namely the one
for $\tau$) does not need to be solved as the internal energy density can
be readily computed by inverting the
relation~\eqref{eq:EOSideal}. Additional details of the formulation we
use for the hydrodynamics equations can be found in~\cite{Font03}.

\begin{figure}[t]
\begin{center}
\includegraphics[width=8.25cm]{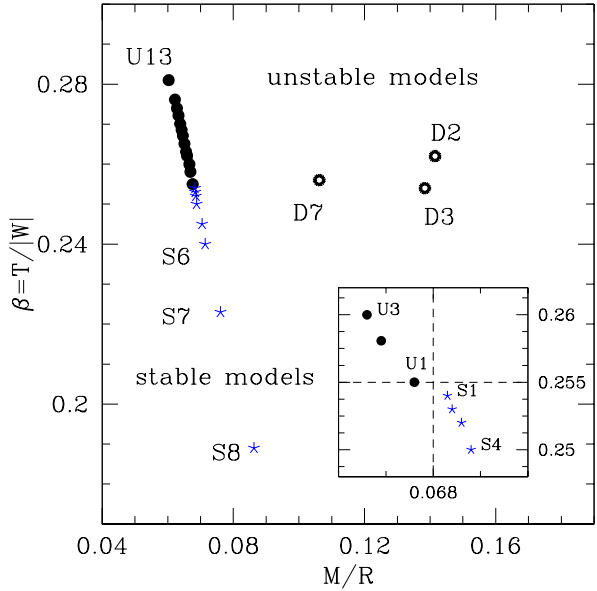}
\end{center}
\vglue-0.5cm
\caption{Position on the $(M/R,\beta)$ plane of the stellar models
  considered. Indicated respectively with stars and filled circles are
  the stable and unstable models belonging to a sequence of constant
  rest mass $M_0 \simeq 1.51\ M_{\odot}$. Open circles refer instead
  to models which do not belong to the sequence, are also unstable and
  were first investigated in ref.~\cite{Shibata00b}. Finally, the
  inset shows a magnification of the region where the threshold of the
  instability has been located for the sequence of models
  investigated.}
\label{fig:beta vs compactness}
\end{figure}

\begin{figure*}[t]
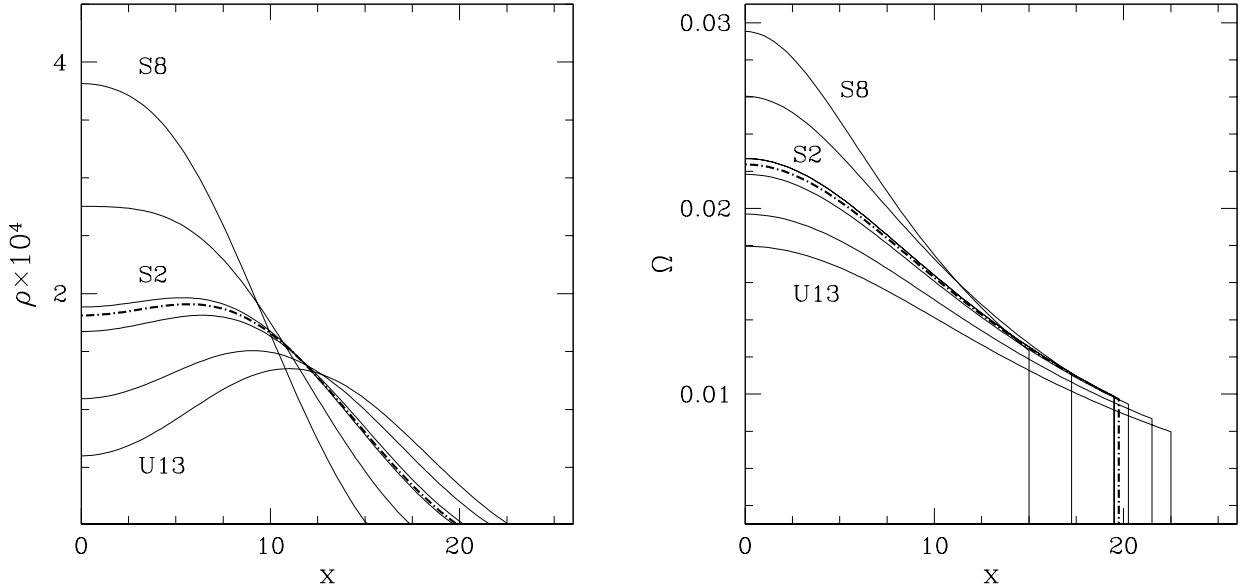

\begin{center}
\vbox{\hbox{ \includegraphics[width=8.25cm]{\imgname{rho_vs_x_0}}
\hskip 0.5cm
\includegraphics[width=8.25cm]{\imgname{omega_vs_x_0}} }}
\end{center}
\vglue-0.8cm
\caption{Initial profiles of the rest-mass density $\rho$ (left panel)
  and of the angular velocity $\Omega$ (right panel) for models
  {S8}, {S7}, {S2}, {U1}, {U3}, {U11} and {U13}. Indicated with a
  dot-dashed line is the profile for the first unstable model ({U1})
  with $\beta$ = 0.255. Note that this is not the first model having an
  off-centered maximum of the rest-mass density.}
\label{fig:InitialProfile}
\end{figure*}

\section{Initial Data}
\label{sec:initial}

The initial data for our simulations are computed as stationary
equilibrium solutions for axisymmetric and rapidly rotating relativistic
stars in polar coordinates~\cite{Stergioulas95}. In generating these
equilibrium models we assumed that the metric describing an axisymmetric
and stationary relativistic star has the form
\begin{equation} 
\begin{split}
ds^{2} = - e^{\mu+\nu} dt^{2}
         + e^{\mu-\nu} r^{2} \sin^{2}\theta (d\phi-\omega dt)^{2}
\\
         +e^{2\xi}(dr^{2}+r^{2} d\theta^{2})
\end{split}
\end{equation} 
where $\mu$, $\nu$, $\omega$ and $\xi$ are space-dependent metric
functions. Similarly, we assumed the matter to be
characterized by a non-uniform angular velocity distribution of the form
\begin{equation} 
\Omega_{c} -\Omega   =   \frac{r_{e}^{2}}{\hat{A}^{2}}
       \left[ \frac{(\Omega-\omega) r^2   \sin^2\theta e^{-2\nu}
              }{1-(\Omega-\omega)^{2} r^2 \sin^2\theta e^{-2\nu}
       }\right] \ ,
\label{eq:velocityProfile}
\end{equation} 
where $r_{e}$ is the coordinate equatorial stellar radius and the 
coefficient $\hat{A}$ is a measure of the degree of
differential rotation, which we set to $\hat{A}=1$ in analogy with other
works in the literature. Once imported onto the Cartesian grid and
throughout the evolution, we compute the angular velocity $\Omega$ (and the 
period $P$) on the $(x,y)$ plane as 
\begin{equation}
\Omega =\frac{u^{\phi}}{u^{0}}
       =\frac{u^{y} \cos \phi - u^{x} \sin \phi}{u^{0}\sqrt{x^2+y^2}}
       \, ,
       \qquad P=\frac{2\pi}{\Omega} 
\end{equation}
and other characteristic quantities of the system, such as the {\it
  baryonic} mass $M_0$, the gravitational mass $M$, the angular
momentum $J$, the rotational kinetic energy ${T}$ and the
gravitational binding energy ${W}$ as
\begin{eqnarray}
\label{relevantquantities}
M   & \equiv & \int\!d^3\!x\, \left( -2T^0_0+T^\mu_\mu \right) \alpha \sqrt{\gamma}
\quad ,  \label{eq:DEF M}\\
M_0 & \equiv & \int\!d^3\!x\,  D  
\quad , \label{eq:DEF M0}\\
{E_{\rm int}} & \equiv & \int\!d^3\!x\, D \epsilon
\quad , \label{eq:DEF Ein}\\
J   & \equiv & \int\!d^3\!x\, T^0_\phi \alpha \sqrt{\gamma}
\quad , \label{eq:DEF J}\\
{T} & \equiv & \frac{1}{2} \int\!d^3\!x\, \Omega  T^0_\phi  \alpha \sqrt{\gamma} 
\quad , \label{eq:DEF T}\\
{W} & \equiv & {T} + {E_{\rm int}} + M_0 - M 
\ , \label{eq:DEF W}
\end{eqnarray}
where $\alpha\, \sqrt{\gamma}$ is the square root of the four-dimensional 
metric determinant. We recall that the definitions of
quantities such as $J$, ${T}$, ${W}$ and $\beta$ are meaningful only
in the case of stationary axisymmetric configurations and should
therefore be treated with care once the rotational symmetry is lost.

All the equilibrium models considered here have been calculated using the
relativistic polytropic EOS (\ref{eq:EOSisentropic}) with $K=100$ and
$\Gamma=2$ and are members of a sequence having a constant amount of
differential rotation with $\hat{A}=1$ and a constant rest mass of $M_0
\simeq 1.51\ M_{\odot}$ (a part of this sequence has also been considered
in refs.~\cite{Stergioulas:2004} and~\cite{Dimmelmeier05} as models
$A8-A10$). These are collected in Fig.~\ref{fig:beta vs compactness} in a
compactness/instability-parameter plot, where we have indicated with
stars the stable models (S1--S8) and with filled circles the unstable
ones (U1--U11). In addition, in order to compare with previous results,
we have also considered three other models (D2, D3, D7), first
investigated in ref.~\cite{Shibata00b}, which have larger masses and
compactnesses. These models are unstable and are marked by circles in
Fig.~\ref{fig:beta vs compactness}. Finally, the inset shows a
magnification of the region where the threshold (indicated with a dashed
line) of the instability has been located for the members of the
sequence.

The main properties of all the considered models are reported in
Table~\ref{table:models}, while we show in
Fig.~\ref{fig:InitialProfile} the profiles of the rest-mass density $\rho$ (left
panel) and of the rotational angular velocity $\Omega$ (right panel)
for some of the models in the constant--rest-mass
sequence. Note that the position of the maximum of the rest-mass
density coincides with the center of the star only for models with low
$\beta$; for those with a larger $\beta$, the maximum of the rest-mass
density resides, instead, on a circle on the equatorial plane.
Finally, indicated with a dot-dashed line in
Fig.~\ref{fig:InitialProfile} is the profile for the first unstable
model ({U1}) with $\beta = 0.255$. Note that this is not the first
model having an off-centered maximum of the rest-mass density.

\begin{table*}
\newcommand{\SKIP}{{~~~~~~~~~~~~~~}}
\newcommand{\comp}{{$M/R_e$}}
\newcommand{\rhoc}{{$\rho_{c}$}}
\newcommand{\Rrate}{{$r_{p}/r_{e}$}}
\newcommand{\msec}{{\small (ms)}}
\newcommand{\expq}{~~{$\scriptstyle (10^{-4})$}~~}
\newcommand{\expD}{~~{$\scriptstyle (10^{-2})$}~~}
\begin{tabular}{|cc|cc|cccc|ccc|ccc|}
\hline\hline 
Model &  &\rhoc~~ &\Rrate   &$R_{e}$&$M_{0}$&  $M$  & \comp  &  $J$  &$P_{a}$&$P_{e}$& $T$     &  $W$    & $\beta$ \\
      &  & \expq  &\SKIP    &\SKIP  &\SKIP  &\SKIP  & \SKIP  &\SKIP  & \msec & \msec & \expD & \expD & \SKIP   \\ 
\hline   
U13   &  & 0.5990 & 0.20010 & 24.31 & 1.505 & 1.462 & 0.0601 & 3.747 & 1.723 & 3.910 & 2.183 & 7.764 & 0.2812  \\
U12   &  & 0.9940 & 0.24150 & 23.52 & 1.508 & 1.462 & 0.0622 & 3.591 & 1.599 & 3.654 & 2.272 & 8.228 & 0.2761  \\
U11   &  & 1.0920 & 0.25010 & 23.31 & 1.507 & 1.460 & 0.0627 & 3.541 & 1.572 & 3.597 & 2.284 & 8.327 & 0.2743  \\
U10   &  & 1.1960 & 0.25860 & 23.08 & 1.508 & 1.460 & 0.0633 & 3.496 & 1.542 & 3.536 & 2.302 & 8.461 & 0.2721  \\
U9    &  & 1.2840 & 0.26550 & 22.88 & 1.508 & 1.460 & 0.0638 & 3.457 & 1.517 & 3.486 & 2.316 & 8.575 & 0.2701  \\
U8    &  & 1.3470 & 0.27030 & 22.73 & 1.508 & 1.460 & 0.0642 & 3.428 & 1.500 & 3.450 & 2.325 & 8.659 & 0.2686  \\
U7    &  & 1.4060 & 0.27470 & 22.59 & 1.509 & 1.460 & 0.0647 & 3.402 & 1.484 & 3.417 & 2.334 & 8.741 & 0.2671  \\
U6    &  & 1.4810 & 0.28030 & 22.40 & 1.508 & 1.459 & 0.0651 & 3.363 & 1.465 & 3.377 & 2.341 & 8.832 & 0.2651  \\
U5    &  & 1.5530 & 0.28560 & 22.22 & 1.508 & 1.458 & 0.0656 & 3.326 & 1.446 & 3.339 & 2.346 & 8.920 & 0.2631  \\
U4    &  & 1.5880 & 0.28810 & 22.13 & 1.508 & 1.458 & 0.0659 & 3.310 & 1.437 & 3.321 & 2.351 & 8.970 & 0.2621  \\
U3    &  & 1.6720 & 0.29430 & 21.92 & 1.506 & 1.456 & 0.0664 & 3.261 & 1.417 & 3.279 & 2.352 & 9.061 & 0.2596  \\
U2    &  & 1.7230 & 0.29780 & 21.78 & 1.508 & 1.457 & 0.0669 & 3.241 & 1.404 & 3.251 & 2.360 & 9.146 & 0.2581  \\
U1    &  & 1.8120 & 0.30500 & 21.54 & 1.499 & 1.448 & 0.0672 & 3.164 & 1.386 & 3.214 & 2.336 & 9.167 & 0.2549  \\
\hline                                                                                                                                           
S1    &  & 1.8600 & 0.30700 & 21.42 & 1.512 & 1.460 & 0.0682 & 3.191 & 1.368 & 3.180 & 2.384 & 9.388 & 0.2540  \\
S2    &  & 1.8850 & 0.30900 & 21.35 & 1.510 & 1.458 & 0.0683 & 3.170 & 1.364 & 3.170 & 2.378 & 9.396 & 0.2531  \\
S3    &  & 1.9160 & 0.31100 & 21.27 & 1.512 & 1.459 & 0.0686 & 3.160 & 1.356 & 3.153 & 2.385 & 9.458 & 0.2522  \\
S4    &  & 1.9620 & 0.31500 & 21.14 & 1.504 & 1.452 & 0.0687 & 3.111 & 1.348 & 3.137 & 2.363 & 9.439 & 0.2503  \\
S5    &  & 2.1280 & 0.32600 & 20.70 & 1.510 & 1.456 & 0.0703 & 3.050 & 1.308 & 3.057 & 2.386 & 9.736 & 0.2451  \\
S6    &  & 2.2610 & 0.33600 & 20.32 & 1.505 & 1.449 & 0.0713 & 2.965 & 1.282 & 3.002 & 2.369 & 9.859 & 0.2403  \\
S7    &  & 2.7540 & 0.37040 & 19.03 & 1.506 & 1.447 & 0.0760 & 2.741 & 1.189 & 2.812 & 2.360 & 10.56 & 0.2234  \\
S8    &  & 3.8150 & 0.44370 & 16.70 & 1.506 & 1.439 & 0.0862 & 2.322 & 1.048 & 2.531 & 2.255 & 11.96 & 0.1886  \\
\hline                                                                                                                                             
\hline                                                                                                                                             
D2    &  & 3.1540 & 0.27500 & 18.29 & 2.771 & 2.587 & 0.1414 & 7.620 & 0.735 & 2.051 & 9.256 & 35.28 & 0.2624  \\
D3    &  & 3.7250 & 0.30000 & 17.85 & 2.640 & 2.466 & 0.1382 & 6.827 & 0.731 & 2.026 & 8.450 & 33.11 & 0.2544  \\
D7    &  & 2.7960 & 0.30000 & 19.56 & 2.188 & 2.075 & 0.1061 & 5.386 & 0.959 & 2.442 & 5.390 & 21.06 & 0.2561  \\
\hline 
\hline 
\end{tabular}
\caption{ Main properties of the stellar models used in the
  simulations. Starting from the left the different columns report:
  the central rest-mass density $\rho_c$, the ratio between the polar
  and the equatorial coordinate radii $r_{p}/r_{e}$, the proper
  equatorial radius $R_{e}$, the rest mass $M_{0}$, the gravitational
  mass $M$, the compactness $M/R_e$, the total angular momentum $J$,
  the rotational periods at the axis $P_{a}$ and at the equator
  $P_{e}$, the rotational energy $T$ and the binding energy $W$, and
  their ratio $\beta$ (instability parameter).}
\label{table:models}
\end{table*}

As mentioned in the Introduction, numerical simulations of the
dynamical bar-mode instability have traditionally been sped up by
introducing sometimes very large initial $m$=2 deformations. The
rationale behind this is simple since a seed perturbation has the
effect of reducing the time needed for the instability to develop and
thus the computational costs. However, as we will discuss in detail in
Section \ref{sec:effectPERT}, the introduction of any perturbation
(especially when this is not a small one) may lead to spurious effects
and erroneous interpretations. Although in almost all of our
simulation we have evolved purely equilibrium models and simply used
the truncation errors to trigger the instability, we have also
considered models which are initially perturbed so as to determine the
effect of these perturbations on the evolution of the instability. In
these cases, we have modified the equilibrium rest-mass density
$\rho_0$ with a perturbation of the type
\begin{equation}
\label{eq:densPert}
\delta\rho_2(x,y,z) = \delta_{2}  \rho_0
	\frac{x^{2}-y^{2}}{r_{e}^{2}} \ ,
\end{equation}
where $\delta_2$ is the magnitude of the $m$=2 perturbation (which we
usually set to be $\delta_2\simeq 0.01-0.3$). This perturbation has
then the effect of superimposing on the axially symmetric initial
model a bar-mode deformation that is much larger than the (unavoidable)
$m$=4-mode perturbation introduced by the Cartesian grid
discretization. In addition to a bar-mode deformation and in order to
test the effect of a pre-existing $m$=1-mode perturbation we also used
in some cases ({\it cf.} Section \ref{sec:effectPERT}) an $m$=1-mode
density perturbation of the type
\begin{equation}
\label{eq:densPert1}
\delta\rho_1(x,y,z) = \delta_{1} 
	\sin \left( \phi + \frac{2 \pi \varpi }{r_{e}} \right) \rho_0 \ ,
\end{equation}
with $\delta_1 = 0.01$. Finally, after the addition of
a perturbation of the type~(\ref{eq:densPert}) or
(\ref{eq:densPert1}), we have re-solved the Hamiltonian and momentum
constraint equations, in order to enforce that the
initial constraint violation is at the truncation-error level.

\section{Methodology of the Analysis }
\label{sec:methodology}

A number of different quantities are calculated during the evolution
to monitor the dynamics of the instability. Among them is the
quadrupole moment of the matter distribution, which we compute in
terms of the conserved density $D$ rather than of the rest-mass
density $\rho$ or of the $T_{00}$ component of the stress energy
momentum tensor
\begin{equation}
\label{eq:defQuadrupole}
I^{jk} = \int\! d^{3}\!x \; D \; x^{j} x^{k} \ .
\end{equation}
Of course, the use of $D$ in place of $\rho$ or of $T_{00}$ is arbitrary
and all the three expressions have the same Newtonian
limit. However, we prefer the form (\ref{eq:defQuadrupole}) because $D$
is a quantity whose conservation is guaranteed by the form chosen for the
hydrodynamics equations (\ref{eq:consform1}). The time variation of
(\ref{eq:defQuadrupole}) (or, rather, suitable combinations of its 
second time derivatives)
will then be used in Section \ref{sec:GW} to characterize the
gravitational-wave emission from the instability.

Once the quadrupole moment distribution is known, the presence of a
bar and its size may be usefully quantified in terms of the distortion
parameters~\cite{Saijo:2001}
\begin{eqnarray}
\eta_{+}      &=\dfrac{{I}^{xx}-{I}^{yy}}{{I}^{xx}+{I}^{yy}} 
\quad , \label{etap}\\
\eta_{\times} &=\dfrac{2\; {I}^{xy}}{{I}^{xx}+{I}^{yy}}  
\quad , \label{etac} \\
\eta          &=\sqrt{\eta_{+}^{2}+\eta_{\times}^{2}} 
\ . \label{eq:Qdistortion} 
\end{eqnarray}
In addition, the quantity (\ref{etap}) can be conveniently used to
quantify both the growth time $\tau_{_{\rm B}}$ of the instability and the
oscillation frequency $f_{_{\rm B}}$ of the unstable bar once the instability
is fully developed. In practice, we perform a nonlinear least-square
fit of the computed distortion $\eta_{+}(t)$ with the trial function
\begin{equation}  
\eta_{+}(t) = \eta_{0} \; e^{t/\tau_{_{\rm B}}} 
                     \cos(2\pi\, f_{_{\rm B}} \, t+\phi_{0}) 
\ .
\label{eq:etafit}
\end{equation}  
Note that all quantities (\ref{etap})--(\ref{eq:Qdistortion}) are
expressed in terms of the coordinate time $t$ and do not represent
therefore invariant measurements at spatial infinity. However, for the
simulations reported here, the lengthscale of variation of the lapse
function at any given time is always larger than twice the stellar
radius at that time, ensuring that the events on the same timeslice
are also close in proper time.

Unless stated differently, we generally do not impose any boundary
condition enforcing certain symmetries. As a result, during the
evolution the compact star is not constrained to be centered at the
origin of the coordinate system and, in order to monitor the relative
motion of the rest-mass density distribution with respect to
the coordinate system, we compute the first momentum of the rest-mass
density distribution
\begin{eqnarray}
X_{cm}^i&=& \displaystyle  \frac{1}{\widetilde{M}} 
	\; \int\!\! d^3\!x \; \rho \;  x^i \ ,
\label{eq:def c.o.m.}
\end{eqnarray}
where $\widetilde{M}\equiv\int\! d^3\!x \; \rho$. These
quantities are reminiscent of the Newtonian definition of the centre
of mass of the star but, because they are not gauge-invariant
quantities, they are not expected to be constant during the
evolution. However, since in a Newtonian framework a time-variation
of one of the $X^{i}_{\rm cm}$ would signal a nonzero momentum in that
direction, we monitor these quantities as a measure of the overall
accuracy of the simulations. Note also that, since the concept of the centre
of mass is well defined in a Newtonian context only, equivalent
definitions to (\ref{eq:def c.o.m.}) could be made in terms of $D$ or
of $T_{00}$. We have verified that in our simulations no significant
quantitative differences are present among the possible alternative
definitions.

\begin{figure}[t]
\begin{center}
\includegraphics[width=8.25truecm]{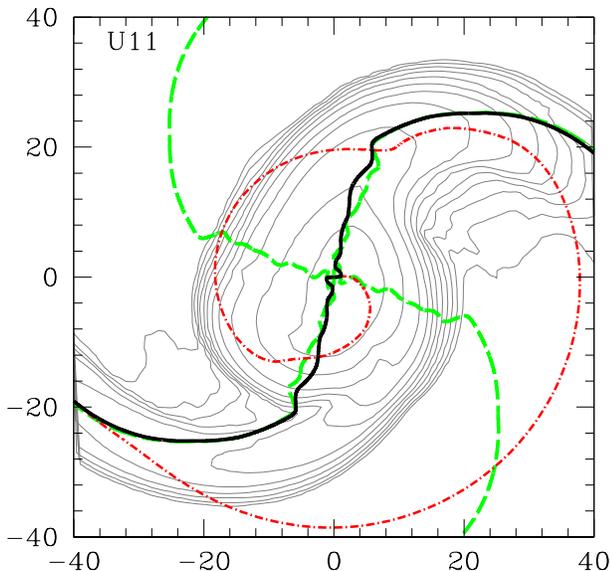}
\end{center}
\vspace{-1cm}
\caption{Mode-phases (solid line for the $m$=2 mode, dashed for the
  $m$=4 mode and dot-dashed for the $m$=1 mode) at different values of
  $\varpi$ overlapped with isocontours of the rest-mass density for model
  {U11} at 25.7ms.}
\label{fig:spiralU11}
\end{figure}

In addition, as a fundamental tool to describe \textit{and understand}
the nonlinear properties of the development and saturation of the
instability, we decompose the rest-mass density into its Fourier modes
so that the ``power'' of the $m$-th mode is defined as
\begin{equation}
{P}_m \equiv\int\!d^3\!x  \, \rho \, e^{{\rm i} m \phi} 
\label{eq:modes}
\end{equation}
and the ``phase'' of the $m$-th mode is defined as
\begin{equation}
{\phi}_m \equiv \arg ({P}_m) 
\ . \label{eq:phimodes}
\end{equation}
The phase ${\phi}_m$ essentially provides the instantaneous
orientation of the $m$-th mode when this has a nonzero power and is
expected to have a harmonic time dependence when the corresponding
mode has a fully developed mode-component.

An important clarification to make is that, despite their
denomination, the Fourier modes (\ref{eq:modes}) do not represent
proper eigenmodes of oscillation of the star. While, in fact, the
latter are well defined only within a perturbative regime, the former
simply represent a tool to quantify, within a fully nonlinear regime,
what are the main components of the rest-mass distribution. Stated
differently, we do not expect that quasi-normal modes of oscillations
are present but in the initial and final stages of the instability,
for which a perturbative description is adequate.

Note also that the diagnostic quantities (\ref{eq:modes}) are closely
related to the \textit{``dipole diagnostic''} $D={P}_1/M$ and {\it
``quadrupole diagnostic''} $Q={P}_2/M$ of ref.~\cite{Saijo:2003}. For
some selected models we have restricted the integration domain in
eqs.~(\ref{eq:modes}) and (\ref{eq:phimodes}) to the equatorial
[\textit{i.e.} $(x,y)$] plane and performed an integration in the
azimuthal angle $\phi$ only. In this way the corresponding quantities
\begin{eqnarray}
{\tilde P}_m(\varpi) 
&\equiv& \int_{z=0} \!\!\!d\phi\; 
	\rho (\varpi\cos(\phi),\varpi\sin(\phi)) 
        e^{{\rm i} m \phi}  
\ , \\[2mm]
{\tilde \phi}_m(\varpi) 
&\equiv& \arg ({\tilde P}_m(\varpi)) 
\ , \label{eq:tildemodesR}
\end{eqnarray}
have an explicit dependence on the cylindrical radial coordinate $\varpi$
only. The quantities (\ref{eq:tildemodesR}) have the advantage that they
can be used to check the \textit{``coherence of the mode''} since
${\tilde \phi}_m(\varpi)$ should be independent of $\varpi$ when the
$m$-th mode is a global property of the matter distribution. As an
example we show in Fig.~\ref{fig:spiralU11} the phases for the $m$=1, 2
and $m$=4 modes for model U11 when the bar is still fully developed, just
before the bar loses its coherence. Note that the $m$=1 mode shows a
spiral-like pattern inside the star, while both ${\tilde \phi}_2$ and
${\tilde \phi}_4$ acquire a radial dependence in the outer parts of the
star, where the bar deformation is absent. A similar behaviour for the
${\tilde \phi}_m(\varpi)$ has been observed in all the performed
simulations.

\section{General features of the dynamics}
\label{sec:General Dynamics}

\begin{figure*}[t]
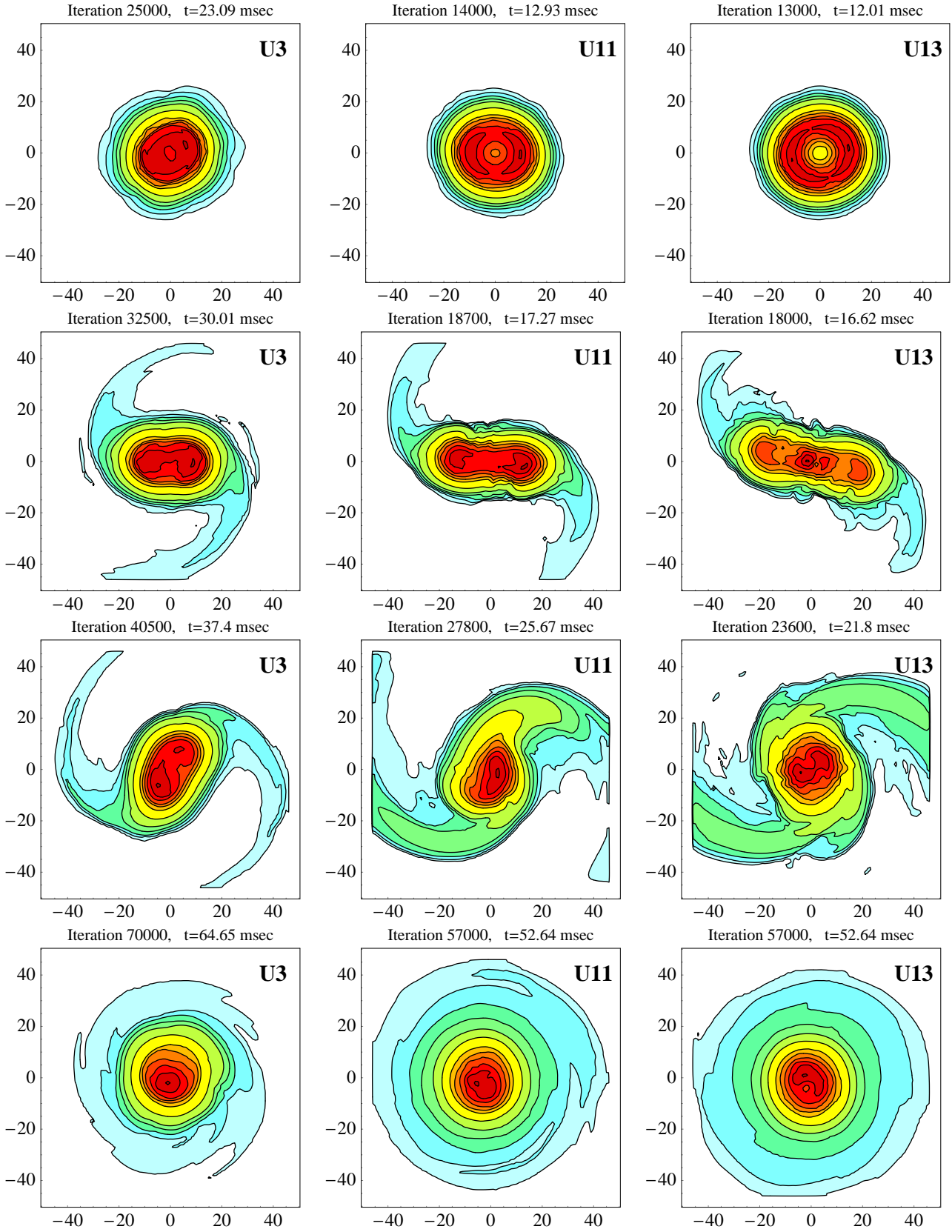

\begin{center}
\vskip -1.0cm
\vbox{\hbox{
\includegraphics[width=5.5cm]{\COLORimgname{snapshot_A11_rho_0}} \hskip 0.25cm
\includegraphics[width=5.5cm]{\COLORimgname{snapshot_A12_rho_0}} \hskip 0.25cm
\includegraphics[width=5.5cm]{\COLORimgname{snapshot_A13_rho_0}} }}
\vskip +0.0cm
\vbox{\hbox{
\includegraphics[width=5.5cm]{\COLORimgname{snapshot_A11_rho_1}}\hskip 0.25cm
\includegraphics[width=5.5cm]{\COLORimgname{snapshot_A12_rho_1}}\hskip 0.25cm
\includegraphics[width=5.5cm]{\COLORimgname{snapshot_A13_rho_1}} }}
\vskip +0.0cm
\vbox{\hbox{
\includegraphics[width=5.5cm]{\COLORimgname{snapshot_A11_rho_2}}\hskip 0.25cm
\includegraphics[width=5.5cm]{\COLORimgname{snapshot_A12_rho_2}}\hskip 0.25cm
\includegraphics[width=5.5cm]{\COLORimgname{snapshot_A13_rho_2}} }}
\vskip +0.0cm
\vbox{\hbox{
\includegraphics[width=5.5cm]{\COLORimgname{snapshot_A11_rho_3}}\hskip 0.25cm
\includegraphics[width=5.5cm]{\COLORimgname{snapshot_A12_rho_3}}\hskip 0.25cm
\includegraphics[width=5.5cm]{\COLORimgname{snapshot_A13_rho_3}} }}
\vskip -1.0cm
\end{center}
\caption{Snapshots of the evolution of models {U3}, {U11} and {U13} at
  various times.  The different columns refer to the three models and
  show isodensity contours for $\rho= 0.9,0.8,0.7,0.6,0.5\time 2^{-2j}
  \times \rho_{\rm max}$, where $j=1,\ldots,6$ and $\rho_{\rm max}$ is
  the maximum value of $\rho$ in each panel.  The above models were
  evolved on a $193\times193\times68$ grid with grid coordinate
  resolution of 0.5 $M_\odot$ (0.74km) and imposing equatorial
  symmetry. The time evolution of some quantities characterising
  these models is reported in Figs.~\ref{fig:GeneralQuadrupoleU3i52},
  \ref{fig:GeneralQuadrupoleU11i36} and
  \ref{fig:GeneralQuadrupoleU13i53}.  }
\label{fig:snapshot}
\end{figure*}

\begin{figure}[t]
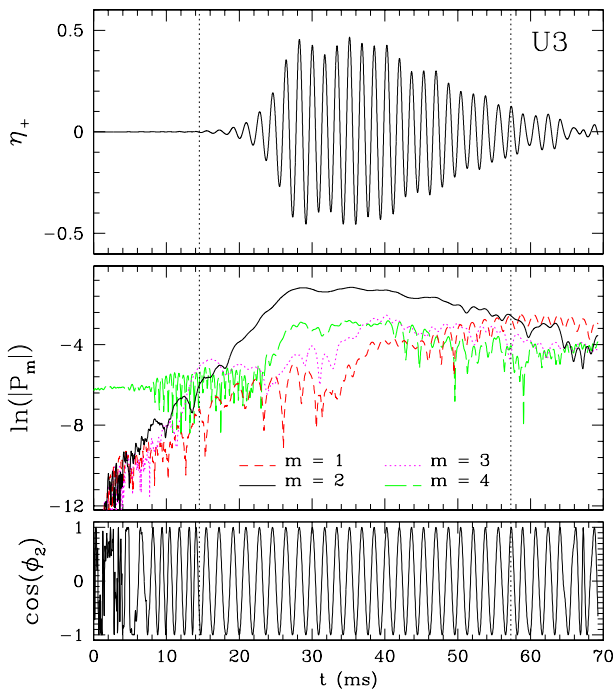

\begin{center}
\mbox{\includegraphics[width=8.6truecm]{\imgname{eta52}}}\\
\mbox{\includegraphics[width=8.6truecm]{\imgname{md52}}}\\
\mbox{\includegraphics[width=8.6truecm]{\imgname{fs52}}}\\
\vspace{-0.7cm}
\end{center}
\caption{Time evolution of the instability for model {U3}. The top
  panel shows the behaviour of the quadrupole distortion parameter
  $\eta_+$ [\textit{cf.} eq.~(\ref{eq:Qdistortion})], the middle panel
  reports the behaviour of the power in the Fourier modes $m$=1, $2$, $3$
  and $4$, while the bottom panel displays the phase of the $m$=2 mode. }
\label{fig:GeneralQuadrupoleU3i52}
\end{figure}

\begin{figure}[t]
\begin{center}
\mbox{\includegraphics[width=8.6truecm]{\imgname{eta36}}}\\
\mbox{\includegraphics[width=8.6truecm]{\imgname{md36}}}\\
\mbox{\includegraphics[width=8.6truecm]{\imgname{fs36}}}\\
\end{center}
\vspace{-0.7cm}
\caption{The same as Fig.~\ref{fig:GeneralQuadrupoleU3i52} but for
  model {U11}.}
\label{fig:GeneralQuadrupoleU11i36}
\end{figure}

\subsection{The tests on stable models}
\label{sec:stable}

Before investigating the nonlinear dynamics of unstable stellar
models, we have carried out a systematic investigation of the ability
of our code to perform long-term stable and accurate evolutions of
stable stellar models. In particular, we have considered the time
evolution of two of the differentially rotating models discussed in
ref.~\cite{Stergioulas:2004,Dimmelmeier05}, namely models S7 and S8,
and have followed their dynamics for 24 and 35 axial rotation periods,
respectively. In both cases the stellar models remain stable and the
density and velocity fluctuations in the stellar interior are smaller
than 2\% during the whole simulation. This is a rather remarkable
result in fully 3D simulations and it is worth stressing that the
simulations reported in ref.~\cite{Font02c} were not able to go beyond
3 orbital periods for similar values of the grid size and spacing (we
recall that in~\cite{Font02c} a second-order TVD method with the MC
limiter was used in place of the third-order PPM method used here).

In addition, for a more quantitative check of the accuracy of our
simulations, we have computed the frequency of the $f$-mode using the
normalized power spectrum (Lomb's method~\cite{lomb76}) of the coordinate
time evolution of the central rest-mass density.  The calculated values
of 791 Hz for model S8 ({A9}) and of 674 Hz for model S7 ({A10}) are in
very good agreement, with the values of 809 Hz and 685 Hz reported in
ref.~\cite{Dimmelmeier05} and computed using a 2D grid in spherical
coordinates but in the conformally flat approximation of General
Relativity.

\subsection{Common feature of unstable models}
\label{sec:unstable}

In this subsection, we discuss some of the general features of the
dynamics of unstable models, postponing to the following sections
the discussion of more detailed aspects of the instability. Here we 
will focus in particular on the dynamics of three
representative unstable models, namely {U3}, {U11} and {U13}, which
have been selected so that their increasing values for the $\beta$
parameter cover the whole range of interest. For these simulations, we
have used a spatial resolution $\Delta x=0.5\,M_{\odot}$ and a grid of
$193\times193\times68$ cells and imposed a reflection symmetry with
respect the $(x,y)$ plane. As a result, between 80 and 90
gridpoints cover the stars along the $x$ and $y$ axes at time $t=0$.
Note that all the simulations reported here make use of a uniform
grid with the location of the outer boundary being rather close to the
stellar surface; this makes the extraction of gravitational waves
difficult and accounts for a very small but nonzero loss of mass and
angular momentum (because of matter escaping the computational box).

In Fig.~\ref{fig:snapshot} we show some representative snapshots of
the rest-mass density at four different times for three different
unstable models (one column for each model). In particular, each row
refers to one of the four representative stages in which the dynamics
of the bar can be divided. These are: \textit{(a)} exponential growth
of the $m$=2 mode and $m$=3 mode (first row); \textit{(b)} saturation
of the instability, development of spiral arms and progressive
attenuation of the bar deformation (second row); \textit{(c)} crossing
of $m$=3 mode and $m$=4 mode and consequent attenuation of the bar,
emergence of the $m$=1 mode as the dominant one (third row);
\textit{(d)} suppression of the {bar deformation} and emergence of an almost
axisymmetric configuration (fourth row). Note that while these stages
are present in all these three models, the coordinate times at which
they take place (indicated in the upper part of each panel), as well
as the amplitude of the deformation, depend on the parameters defining
the initial models, most notably $\beta$ and $M_0$.

Understanding the occurrence of these four stages during the onset,
development and suppression of the {bar deformation} represents our effort
to go beyond the standard phenomenological discussion of the nonlinear
dynamics of the instability often encountered in the literature. An
important tool in this discussion will be offered by the time
evolution of the Fourier mode-decomposition (\ref{eq:modes}) discussed
in Sec.~\ref{sec:methodology}. As we will show below, relating the
evolution of these quantities to the evolution of the mode phases
$\phi_m$ and to the changes in the deformation of the star $\eta_{+},
\eta_{\times}$ will allow us to provide a consistent description of
the four stages of the instability.

\begin{figure}[t]
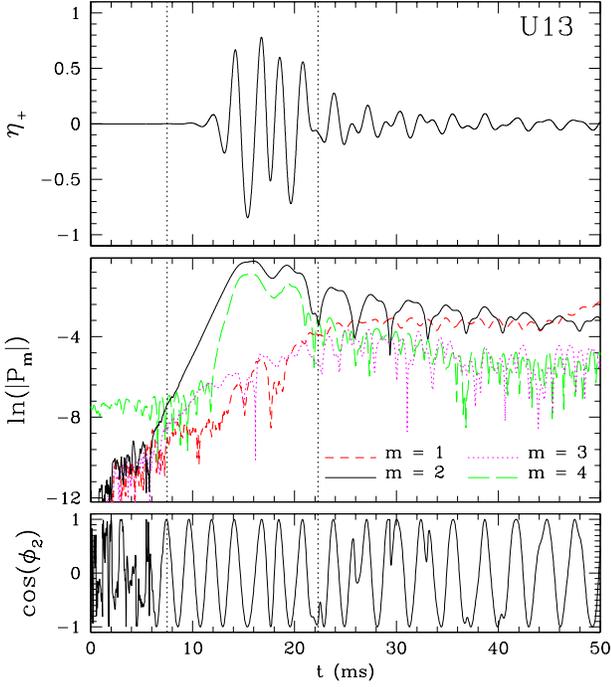

\begin{center}
\mbox{\includegraphics[width=8.6truecm]{\imgname{eta53}}}\\
\mbox{\includegraphics[width=8.6truecm]{\imgname{md53}}}\\
\mbox{\includegraphics[width=8.6truecm]{\imgname{fs53}}}\\
\end{center}
\vspace{-0.7cm}
\caption{The same as Fig.~\ref{fig:GeneralQuadrupoleU3i52} but for
  model {U13}.}
\label{fig:GeneralQuadrupoleU13i53}
\end{figure}

We start our discussion by reporting in
Figs.~\ref{fig:GeneralQuadrupoleU3i52},
~\ref{fig:GeneralQuadrupoleU11i36}
and~\ref{fig:GeneralQuadrupoleU13i53} the history of the instability
for models U3, U11 and U13. Starting from the upper panels, these
figures show: the time evolution of the distortion parameter
$\eta_{+}$ (a very similar behaviour can be shown for the other
distortion parameter $\eta_{\times}$), the power $P_m$ in the first
four $m$-modes and the evolution of the phase of the $m$=2 mode. Note
that at the beginning of the simulation, as a result of the Cartesian
discretization, the $m$=4 mode has the largest power. While this can
be reduced by increasing the resolution, the $m$=4 deformation plays
no major role in the development of the instability, which is soon
dominated by the lower-order modes.

The initial phase of the instability [stage \textit{(a)} in the previous
classification] is clearly characterized by the exponential growth of the
$m$=2 mode and $m$=3 mode, the latter one having a smaller growth rate. A
first interesting mode coupling takes place when the exponentially
growing $m$=3 mode reaches the same power amplitude of the $m$=4 mode, at
which point the two modes exchange their dynamics, with the $m$=4 mode
growing exponentially and the $m$=3 mode reaching saturation. At
approximately the same time, the $m$=1 mode also starts to grow
exponentially but with a growth rate which is smaller than that of the
other modes. Note that this ``mode-amplitude crossing'' between the
$m$=3 and $m$=4 modes also signals the time when collective
phenomena start to be fully visible (note that this mode-amplitude
crossing is distinct from the ``avoided-crossing'' observed when studying
mode eigenfrequencies along sequences of stellar models). This is shown
with the first vertical dotted line in
Figs.~\ref{fig:GeneralQuadrupoleU3i52}, \ref{fig:GeneralQuadrupoleU11i36}
and~\ref{fig:GeneralQuadrupoleU13i53}, marking the time when the
distortion parameter starts being appreciably different from zero (upper
panel) and the $m$=2 phase assumes a harmonic time dependence (lower
panel).  This stage continues until the $m$=2 mode reaches its maximum
power and the bar has reached its largest extension. During the following
phase [stage \textit{(b)}] the bar instability has reached a nonlinear
saturation, accompanied by the development of spiral arms which are
responsible for ejecting a small amount of matter and for a progressive
attenuation of the bar extension (see discussion in
Sec.~\ref{subsec:details on models}). Furthermore, when the exponentially
growing $m$=1 mode reaches the same power amplitude of the $m$=3 mode,
the latter, whose growth had slowed down for a while, returns to grow
exponentially.

\begin{figure}[t]
\begin{center}
\mbox{\includegraphics[width=8.25truecm]{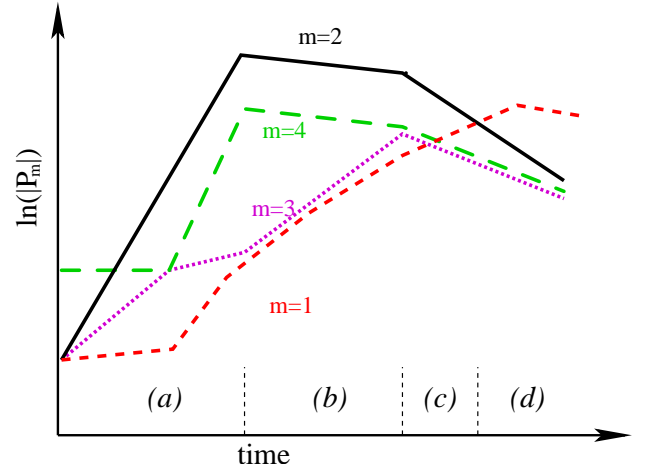}}\\
\vspace{-0.3cm}
\end{center}
\caption{Schematic evolution of the collective modes
  [Eq. \ref{eq:modes}] of the rest-mass density $\rho$. In this
  diagram the instability is distinguished in four representative
  stages: \textit{(a)} exponential growth of the $m$=2 mode and
  $m$=3 mode; \textit{(b)} saturation of the instability, development
  of spiral arms and progressive attenuation of the bar deformation;
  \textit{(c)} crossing of $m$=3 mode and $m$=4 mode and consequent
  attenuation of the bar, emergence of the $m$=1 mode as the dominant
  one; \textit{(d)} suppression of the {bar deformation} and emergence of an
  almost axisymmetric configuration.}
\label{fig:mode_evolution_scheme}
\end{figure}

The following phase of the instability [stage \textit{(c)}] sees modes
$m$=1, 3 and 4 reach comparable powers and this marks the time when the
bar deformation has a sudden decrease. As a result of this crossing among
the three modes, only the $m$=1 mode will continue to grow, while the
$m$=3 and the $m$=4 modes are progressively damped. Finally, stage {\it
(d)} starts when the growing $m$=1 mode reaches power amplitudes
comparable with those of the $m$=4 mode and the final mode-amplitude crossing
takes place. This marks a distinct loss of the bar deformation and the
emergence of an almost axisymmetric rapidly rotating star. This is shown
with the second vertical dotted line in
Figs.~\ref{fig:GeneralQuadrupoleU11i36}
and~\ref{fig:GeneralQuadrupoleU13i53}, highlighting when the distortion
parameter is significantly reduced (upper panel) and the $m$=2-mode phase
loses its harmonic time dependence (lower panel). A schematic and
qualitative diagram summarizing the evolution of the power in the first
four $m$-modes as discussed above is shown in
Fig.~\ref{fig:mode_evolution_scheme} and can be used as an aid for the
interpretation of the quantities computed in
Figs.~\ref{fig:GeneralQuadrupoleU3i52},
~\ref{fig:GeneralQuadrupoleU11i36} and~\ref{fig:GeneralQuadrupoleU13i53}.

We note that the lack of a perturbative study of this process beyond
the linear regime leaves the origins of this interaction between modes
still unclear. Furthermore, since the growth of the $m$=1 mode is not
clearly exponential, especially for slightly overcritical models
(\textit{cf.} Fig.~\ref{fig:GeneralQuadrupoleU3i52} for model U3),
  we have referred to this process as ``mode coupling'' rather than
  considering it as the evidence of an $m$=1 instability. Additional
  perturbative work in this respect will help clarify this aspect.

The general and common features of the dynamics of the
bar-mode instability as deduced from the numerical simulations can be
summarized as follows:
\begin{itemize}

\item the bar deformation is, in general, not a persistent phenomenon
  but is suppressed rather rapidly and over a timescale which is of
  the order of the dynamical one (see also the following Section for
  an additional discussion on this);

\item nonlinear mode couplings take place during the evolution and
  these allow for the growth of other modes besides the
  fastest-growing 
  $m$=2 mode;

\item the growth of other modes has the overall impact of
  progressively attenuating the $m$=2 mode and, consequently, the bar
  deformation, after the instability has saturated;

\item for slightly supercritical models (\textit{e.g.} U3), when the
  power amplitude of the $m$=1 mode has become comparable with the one in
  the $m$=2 mode, the {bar deformation} is suppressed and the star
  evolves towards an almost axisymmetric configuration;

\item for largely supercritical models (\textit{e.g.} U13), the dynamics
  of the instability are so violent and the stellar model so far from
  equilibrium that the strong bar deformation is lost even in the
  absence of mode-coupling effects (see discussion in
  Sec.~\ref{sec:persistence}).

\end{itemize}

\section{Detailed features of the dynamics}
\label{sec:detaileddynamics}

In this Section we discuss some detailed aspects of the instability,
concentrating our attention on the impact that different values of
$\beta$, different boundary conditions, different values of the
initial perturbations, different EOSs and different grid resolutions
or boundary locations have on the onset and development of the instability.
We note that while many of these different prescriptions do not induce
qualitative changes, some of them do change the initial relative amplitude 
of the different modes [and hence the simulation time
needed for the instability to develop and the orientation of the bar
in the $(x,y)$ plane at a given time during the
instability]. In these cases, in order to make meaningful
comparisons, we remove these offsets by choosing a suitable shift in
time $\Delta t$ and in phase $\Delta\phi$ in such a way that the
distortion parameters of the reference model $\eta_+^{R}$ and of the
new one $\eta_+$ have the maximal overlap and are related as
\begin{equation}
\eta_+^{(R)}(t) \simeq \alpha \eta_+(t+\Delta t) + \beta
\eta_\times(t+\Delta t) \ , 
\label{eq:defSHIFT}
\end{equation}
where $\alpha=\cos(\Delta\phi)$, $\beta=\sin(\Delta\phi)$. 

\subsection{Dependence on $\beta$}
\label{subsec:details on models}

The parameter $\beta$ plays a very important role in determining the
properties of the nonlinear dynamics of the instability both with
regard to the growth rate $\tau_{_{\rm B}}$ and to the duration
$\tau_{_{\rm D}}$ of the saturation stage [stage (b) of
  Fig.~\ref{fig:mode_evolution_scheme}].  While the relation between
$\beta$ and $\tau_{_{\rm B}}$ will be discussed in more detail in
Sec.~\ref{sec:threshold}, we here concentrate on how the dynamics of
the bar, once formed, depend on the degree of overcriticality, {\it
  i.e.} on $\beta - \beta_c$, where $\beta_c$ marks the separation
between stable and unstable models. To illustrate this we will
consider two models which are representative of the whole set
considered in Table~\ref{table:models} and which have very different
values of $\beta$ and consequently very distinct behaviours: the
largely overcritical model {U13} and the slightly overcritical model
{U3}.

Model {U13} has $\beta=0.2812$ and is the most unstable of the studied
models, since equilibrium models with larger $\beta$ cannot be
produced along the constant--rest-mass sequence chosen here. The clearest
feature shown in Fig.~\ref{fig:GeneralQuadrupoleU13i53} for
this model is its very rapid growth rate (almost three times larger
than the one for U3 as reported in Tab.~\ref{table:results}), but also
its very effective suppression of the {bar deformation}.  Indeed, the
evolution is so rapid that it is very little affected either by
initial perturbations or by the imposition of additional symmetries
(see next subsections). In this case, in fact, the crossing between
the $m$=2 mode and $m$=1 mode is less evident and the bar deformation
goes through large variations, as shown by the large oscillations in
$P_2$ after $t \simeq 16$ ms, during which time the bar seems to
disappear and then form again soon after. At about $t \simeq 20$ ms
the bar deformation starts disappearing in coincidence with the
mode-amplitude crossing. Again as a result of the very violent
dynamics, the saturation stage is rather short and the bar is
essentially lost after about 8 ms.

\begin{figure}[t]
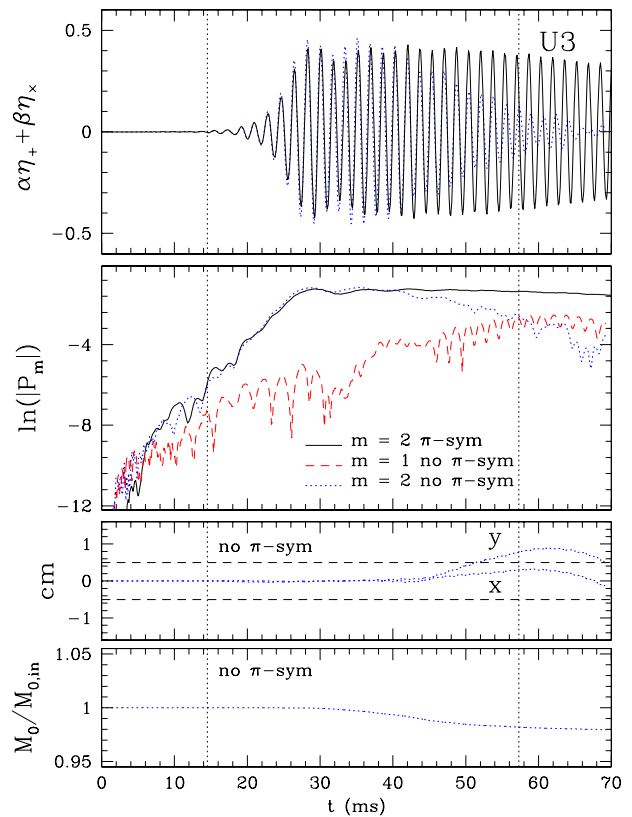

\begin{center}
\mbox{\includegraphics[width=8.6truecm]{\imgname{eta52pi}}}\\
\mbox{\includegraphics[width=8.6truecm]{\imgname{md52pi}}}\\
\mbox{\includegraphics[width=8.6truecm]{\imgname{cm52}}}\\
\mbox{\includegraphics[width=8.6truecm]{\imgname{rm52}}}\\
\end{center}
\vspace{-0.7cm}
\caption{The role of the $\pi$-symmetry on the dynamics for model
  {U3}. Shown from the top are the deformation parameter $\eta$, the
  power in the $m$=2 mode and in the $m$=1 mode (dashed line),
  the evolution of the position of the \mbox{``centre of mass''} (the
  horizontal dashed lines mark the edges of the central cell) and that
  of the rest mass.  The dotted and continuous lines refer to simulations without
  and with $\pi$-symmetry, respectively.}
\label{fig:U3pisym}
\end{figure}

Model {U3}, on the other hand, has $\beta=0.2596$ and shows 
dynamics which are in many respects the opposite of the ones discussed
for model U13.  As shown in Fig.~\ref{fig:GeneralQuadrupoleU3i52}, the
mode evolution is very smooth and, once formed, the bar persists
without significant losses in power.  The growth rate is clearly
smaller and the stage of saturation is much longer (about 30 ms) and
the growth of the $m$=1 mode plays a major role in the damping of the
$m$=2 mode. The transition that leads to the disappearance of the bar
is smooth and it requires many rotation periods. Differently from
model U13, in this case, the properties of the
bar dynamics in the first stage are sensitive to the use of perturbations or to the
imposition of additional symmetries (see next subsections).

\begin{figure}[t]
\begin{center}
\mbox{\includegraphics[width=9.truecm]{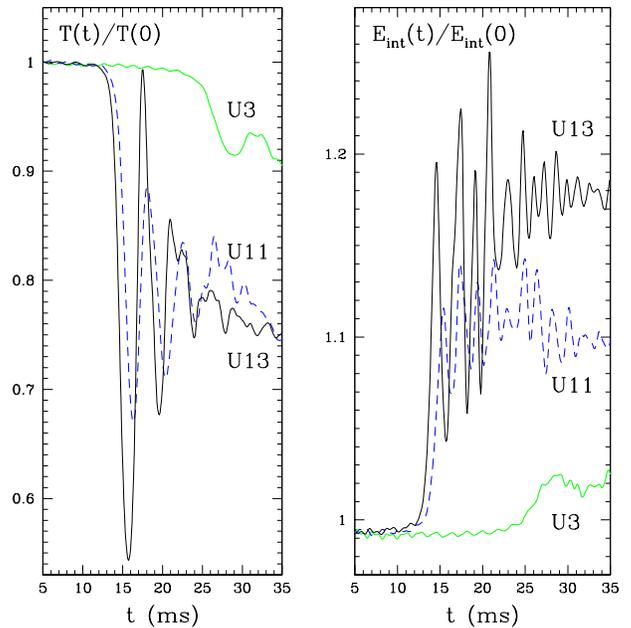}}
\vspace{-1.1cm}
\end{center}
\caption{Dynamics of the rotational kinetic energy $T$ [\textit{cf.}
    eq.~\eqref{eq:DEF T}] and of the internal energy ${E}_{\rm int}$
  [\textit{cf.} eq.~\eqref{eq:DEF Ein}] for models U3, U11 and U13, when
  normalized to their initial values.}
\label{fig:EandT_vs_t}
\end{figure}

Overall, it is reasonable to expect that the persistence of the bar is
strictly related to the degree of overcriticality, with the duration
of the saturation $\tau_{_{\rm D}}$ tending to the radiation-reaction
timescale for a model with $\beta=\beta_c$ and to zero for a model
with $\beta \gg \beta_c$, for which the excess of kinetic rotational
energy may well produce a rapid disruption of the star (see Fig.\
\ref{fig:EandT_vs_t}). The numerical values for $\tau_{_{\rm D}}$ have
been estimated through a nonlinear fit to the evolution of $P_2$ with
3 separated single exponential functions in the three intervals
$[2,t_{a}]$, $[t_{a},t_{b}]$ and $[t_{b},t_{c}]$, where $t_{a}$ and
$t_{b}$ are two free parameters and $t_c$ marks the end of the
simulation.  The estimates for $\tau_{_{\rm D}}$ reported in
Table~\ref{table:results} are still too sparse to be able to delineate
its dependence, beyond the evidence that $\tau_{_{\rm D}} \propto
|\beta - \beta_c|^{-n}$, where $n$ is a positive number. Furthermore,
the reported values have an error of about $1$~ms, as a result of the
used fitting procedure.


\begin{figure}[t]
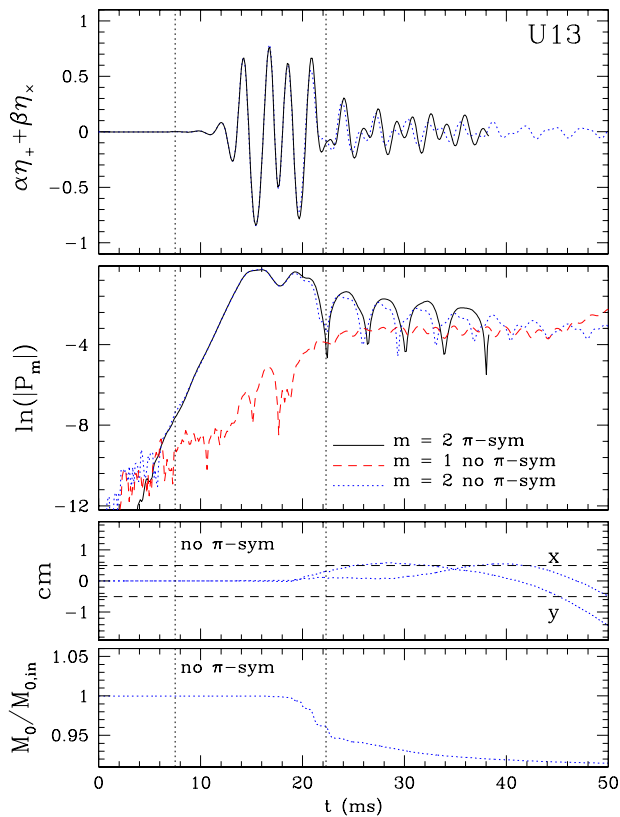

\begin{center}
\mbox{\includegraphics[width=8.6truecm]{\imgname{eta53pi}}}\\
\mbox{\includegraphics[width=8.6truecm]{\imgname{md53pi}}}\\
\mbox{\includegraphics[width=8.6truecm]{\imgname{cm53}}}\\
\mbox{\includegraphics[width=8.6truecm]{\imgname{rm53}}}\\
\end{center}
\vspace{-0.7cm}
\caption{The same as in Fig.~\ref{fig:U3pisym} but for model U13.}
\label{fig:U13pisym}
\end{figure}

\subsection{The role of symmetries}
\label{sec:persistence}

As mentioned in the Introduction, the issue of the persistence of the
bar deformation has been rather controversial
over the years. 

While previous calculations carried out in Newtonian physics and in the
absence of symmetries have highlighted that the {bar deformation} can be
rapidly suppressed as a result of the growth of an $m$=1-mode
deformation~\cite{SmithHouserCantrella96}, subsequent studies have
attributed the growth of the odd mode to inaccurate numerical methods and
supported the idea that the bar should be persistent over a
radiation-reaction timescale and that the use of suitable symmetry
conditions that remove the growth of the odd mode provides a more realistic
description of the bar dynamics~\cite{NewCentrellaTohline00}. In
addition, it has been argued that once an $m$=2-mode perturbation has
developed, only couplings with even modes should be expected and that the
growth of any odd mode should therefore be considered a spurious
numerical artifact.

We believe the above argument not to be valid, except in a linear regime
and in the very idealized case in which it is possible to inject
exclusively an $m$=2 perturbation. In practice, however, any initial
perturbation, either introduced \textit{ad hoc} or by the truncation
error, will excite both even and odd modes and all of these will
couple once a nonlinear regime is reached.

Having said this, it is nevertheless important to verify that the
growth of the $m$=1 mode detected in our simulations is not a
numerical artifact (this is further discussed in Sec.~VI.E) and that
the argument made about the non-persistence of the bar deformation
continues to hold also when boundary conditions with symmetries are
introduced. For this reason, we evolved the models discussed in the
previous Sections also with the use of the so-called $\pi$-symmetry,
ensuring that $f(\varpi,\phi,z) = f(\varpi,\phi+\pi,z)$ for any
variable $f(x^i)$. Clearly, the presence of any odd mode is in this
way impossible by construction. We report in
Figs.~\ref{fig:U3pisym}-\ref{fig:U13pisym} the results of the
simulations for models {U3} and {U13} using this symmetry. The first
two panels from the top show the deformation parameter $\eta(t)$ and
the power in the $m$=2 mode (solid line when the $\pi$-symmetry is
enforced and dotted line otherwise) and in the $m$=1 mode (dashed
line).


\begin{figure*}[ht]
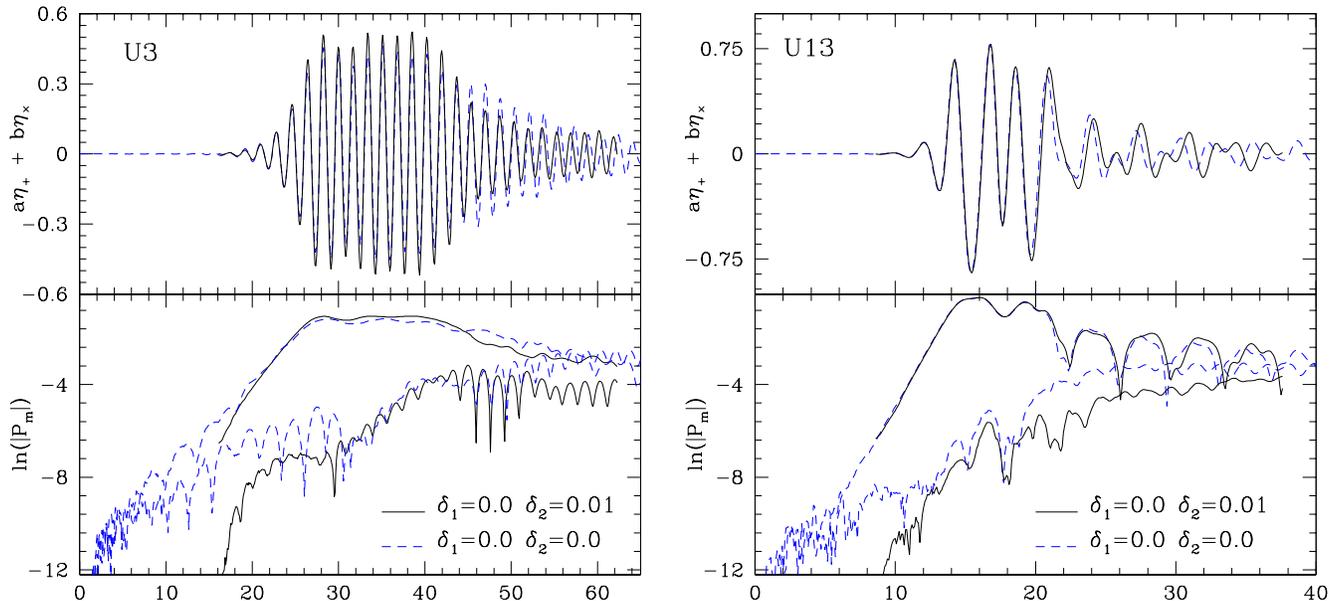

\begin{center}
\includegraphics[width=8.9truecm,height=8.9truecm]{\imgname{pert_a11}}
\includegraphics[width=8.9truecm,height=8.9truecm]{\imgname{pert_a13}}
\end{center}
\vspace{-1cm}
\caption{Effect of an initial $m$=2-mode perturbation on the dynamics of
  the deformation parameter $\eta(t)$ (top sub-panels) and of the modes
  ${P}_2(t)$ and ${P}_1(t)$ (bottom sub-panels) for model {U3} (left) and
  {U13} (right), respectively. The continuous lines represent the
  evolution of the perturbed model after a suitable phase and time
  shifts.}
\label{fig:pert_I}
\end{figure*}

Fig.~\ref{fig:U3pisym} clearly shows that the bar deformation is
essentially persistent in model U3 when the symmetry boundary
conditions are applied, its power amplitude being just slowly
attenuated, mostly because of the entropy production via the
non-isentropic EOS~(\ref{eq:EOSideal}) and a possible small
contribution due to the use of a tenuous atmosphere outside the
star. However, it is important to note that, besides being convergent
and stable, the solution without $\pi$-symmetry is also accurate, as
well as the one without $\pi$-symmetry. This is shown by the second
panel from the bottom in Fig.~\ref{fig:U3pisym}, reporting the
evolution of the position of the ``centre of mass'' as defined in
eq.~(\ref{eq:def c.o.m.}), in the absence of $\pi$-symmetry. The two
horizontal dashed lines in that panel mark the edges of the central
cell and indicate that up to $t\simeq 50$ ms the position of the
centre of mass does not leave the central cell of the grid and that
the exponential growth of the $m$=1 mode (which becomes significant
from $t\simeq 15$ ms) cannot be related to a spurious numerical
effect. After $t\simeq 50$ ms the centre of mass starts to move away
from the centre of the grid and also in this case the motion is not
due to numerical accuracy but rather to the fact that a small amount
of matter (about 2\% of the initial one) is being lost from the grid
as a result of the development of extended spiral arms. This is shown
in the lower panel of Fig.~\ref{fig:U3pisym}, which reports the
evolution of the rest mass when normalized to the initial value and
which clearly shows that the motion of the centre of mass is related
to the loss of rest mass through the grid and thus consistent with the
conservation of linear momentum. As we will further discuss in
Sec.~\ref{sec:effectRES}, the mass loss and the consequent motion of
the centre of mass can be reduced considerably when moving the outer
boundary to larger positions.  As mass leaves the grid, so does
angular momentum, with losses that vary according to the model
considered and ranging from $\sim 3\%$ for model U3 up to $\sim 20\%$
for the more violent model U13. Note, however, that much smaller
angular-momentum losses (\textit{i.e.} less than 1\% for all models)
are in general measured before the mass is shed across the
computational boundaries; as a result, angular momentum can be
conserved to reasonable accuracy by using more
distant outer boundaries (this improvement has been tried
and a discussion on the changes introduced is presented in
Sec.~\ref{sec:effectRES}).

Interestingly, the use of a $\pi$-symmetry does not produce a
significant change in the case of model {U13}. This is true for the
dynamics of the bar (\textit{cf.} the solid and dotted lines in
Fig.~\ref{fig:U13pisym}) and also for the values of $\tau_{_{\rm B}}$ and
$f_{_{\rm B}}$. We believe this is because the dynamics of this
largely overcritical model are not dominated by the mode coupling, but
rather by efficient conversion of rotational kinetic energy into
internal energy. As shown in Fig.~\ref{fig:EandT_vs_t}, which reports
the time evolution of the internal and rotational kinetic energies
when normalized to their initial values, model U13 experiences a
dramatic and rapid increase in the internal energy at the expense of
the kinetic one. This conversion of energy is the largest among the
simulated models and so effective that mode-coupling effects do not
have time to develop. This explains why the use of symmetry conditions
slightly reduces the attenuation of the bar, but cannot prevent its
rapid disappearance.

As a final remark we note that the use of a $\pi$-symmetry produces
only small changes in the values of the bar-pattern frequency
$f_{_{\rm B}}$ or of the growth time $\tau_{_{\rm B}}$ when compared
with the corresponding values computed in the absence of symmetries
(\textit{cf.} Table~\ref{table:results}). This is essentially because
these boundary conditions do not alter the dynamics of the stage of
exponential growth of the bar. They can therefore be used to reduce the
computational costs and pursue the systematic search for the threshold
of the instability discussed in Sec.~\ref{subsec:threshold2}.

\begin{figure*}[t]
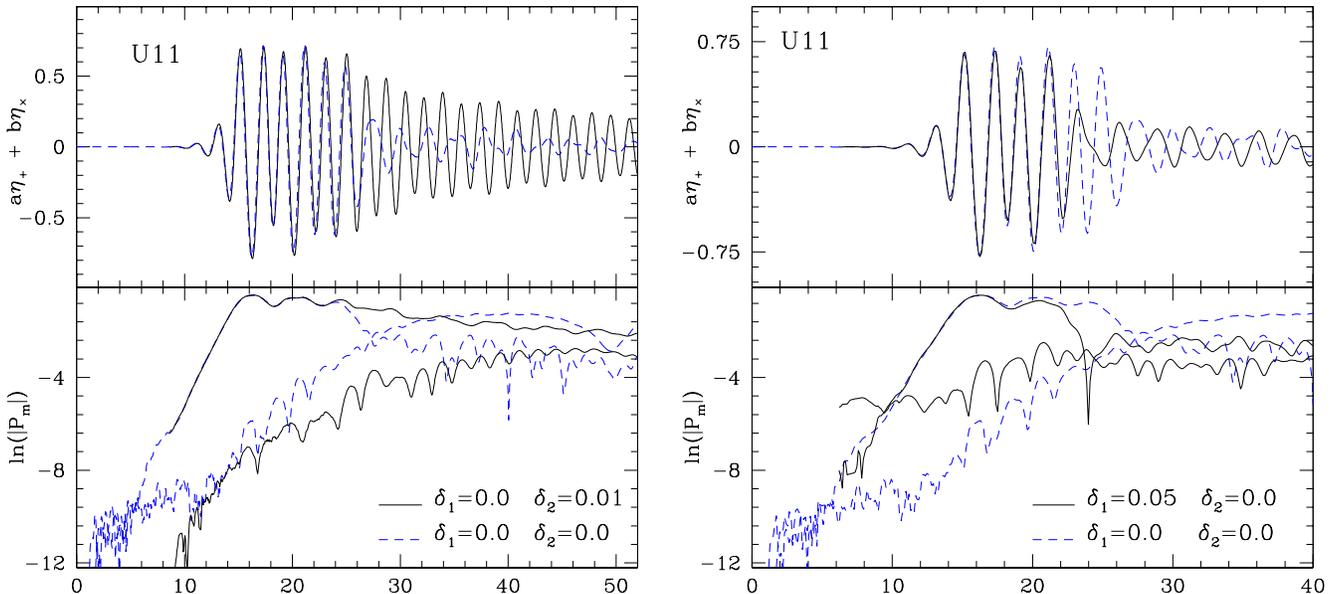

\begin{center}
\includegraphics[width=8.9truecm,height=8.9truecm]{\imgname{pert_a12}}
\includegraphics[width=8.9truecm,height=8.9truecm]{\imgname{pert1_a12}}
\end{center}
\vspace{-1cm}
\caption{Effect of an initial perturbation on the dynamics of the
  deformation parameter $\eta$ (top panels) and of the mode powers
  ${P}_2(t)$, ${P}_1(t)$ for model {U11}. The left panel shows the
  effects of an initial $m$=2-mode perturbation, while the right one
  those of an $m$=1-mode perturbation.}
\label{fig:pert_II}
\end{figure*}

\subsection{The role of the initial perturbation}
\label{sec:effectPERT}

Another common feature of previous works on the bar-mode instability
has been the introduction of a sizeable $m$=2-mode perturbation of the
type shown in eq.~(\ref{eq:densPert}) with the goal of triggering the
instability~\cite{Shibata00b, Saijo2000, Saijo:2001,
  ShibataKarinoEriguchi2002, ShibataKarinoEriguchi2003,
  ShibataSekiguchi2004}. This approach clearly reduces the
computational costs but it is fully justified only when the triggered
mode is the \textit{only unstable} one. However, if other unstable modes
exist, their development may be altered or even suppressed in the
presence of a strong $m$=2-mode perturbation. This is particularly
relevant for the analysis carried out in this work, which has
pointed out that nonlinear mode couplings may trigger the growth of other
modes and significantly modify the dynamics of the instability.

We have therefore considered with care how the introduction of an $m$=2-mode
perturbation influences the onset and the development of the
instability. More specifically, we have added an $m$=2-mode
perturbation of the type shown in eq.~\eqref{eq:densPert} with
$\delta_2=0.01$, solved again the constraint equations and observed
that the impact this has on the development of the instability depends
on the degree of overcriticality. In particular, for models near the
threshold, such as {U3}, the perturbation does not induce changes in
the saturation phase nor in the persistence of the bar, but it does
have the effect of slightly altering the first part of the
evolution, with an increase in the maximum distortion (which at
saturation is $\sim 10 \%$ larger) as well as with an increase in the
growth rate (the growth time $\tau_{_{\rm B}}$ is reduced by $\sim
10~\%$); see the left panel of Fig.~\ref{fig:pert_I} and
Table~\ref{table:results} for a quantitative comparison.

On the other hand, for models that are largely overcritical, such as
U13, and in analogy with what discussed in the previous Section for
the use of symmetric boundary conditions, the introduction of a
perturbation does not have a significant effect and the dynamics are
essentially unaltered (see the right panel of Fig.~\ref{fig:pert_I}).
Finally, for models which are overcritical but not close to the
threshold, such as {U11}, the initial perturbation has a much smaller
impact on both the growth rate and the maximum distortion ({\it
  cf.} Table~\ref{table:results}), but it does increase the duration
of the bar. This is due to the fact that the growth of the $m$=1 mode
is closely related to the one in the $m$=2 mode and its growth can be
delayed and reduced if the latter has initially a non-negligible
power. Because of this, the time at which the two modes have
comparable power will be different and in particular will be postponed
in the perturbed case (see the left panel of Fig.~\ref{fig:pert_II}). Of course, the
converse is also true and modified dynamics for this model are
observed also when an $m$=1-mode perturbation of the type shown in
eq.~\eqref{eq:densPert} is introduced with $\delta_1=0.05$. This is
summarised in the right panel of Fig.~\ref{fig:pert_II}, which shows
that in this case the perturbation reduces the duration of the
saturation stage.

In summary, while the introduction of a seed perturbation (either in
the form of an $m$=1 mode or an $m$=2 mode) does not produce
significant qualitative changes in the dynamics of the instability, it
can result into quantitative changes, most notably in the growth rate,
in the maximum distortion and in the persistence of the bar. The
persistence of the bar, in particular, is enhanced when an $m$=2-mode
perturbation is present. The relevance of these results will need to
be evaluated for those astrophysical scenarios in which long-lasting
bars were simulated, but which were triggered through the introduction
of a perturbation~\cite{ShibataSekiguchi2004,
ShibataKarinoEriguchi2002}.


\begin{figure*}
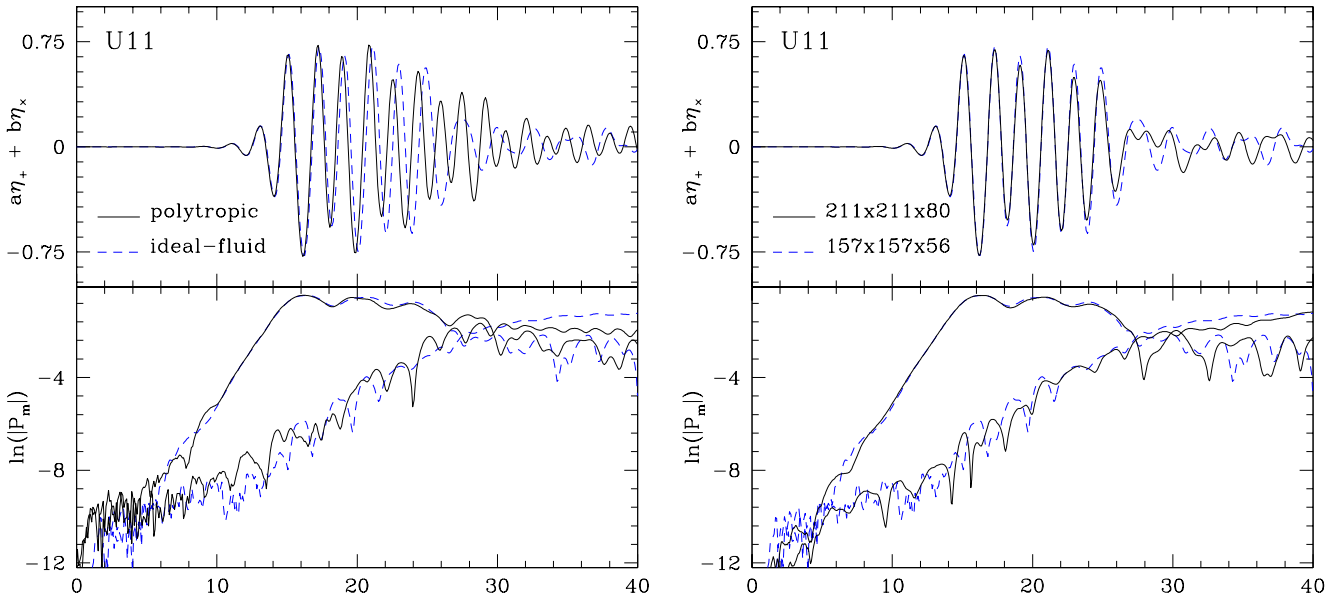

\begin{center}
\includegraphics[width=8.9truecm,height=8.9truecm]{\imgname{eos_a12}}
\includegraphics[width=8.9truecm,height=8.9truecm]{\imgname{border_a12}}
\end{center}
\vspace{-1cm}
\caption{Effects on the evolution of the deformation parameter $\eta$
  (top panels) and the modes ${P}_2(t)$ and ${P}_1(t)$ (bottom panels)
  of model {U11} caused by: (left) the use of the adiabatic polytropic
  EOS; (right) the use of a larger simulation grid, where the distance of
  the outer boundary from the center is increased from 48 $M_\odot$ to 66
  $M_\odot$ on the $(x,y)$-plane and from 32 $M_\odot$ to 47 $M_\odot$
  along the $z$ direction.}
\label{fig:U11adia}
\label{fig:U11_cf}
\end{figure*}

\subsection{The role of the EOS}
\label{sec:effectEOS}

Besides nonlinear mode coupling, another process that could in
principle limit the persistence of the bar is the formation of shocks
(either macroscopical or on smaller scales) that would convert the
excess kinetic energy into internal one. In order to assess the
importance of this process we have compared the evolution of the
relevant unstable models when these are evolved using the
non-isentropic EOS~(\ref{eq:EOSideal}) and when using the (isentropic)
polytropic EOS~(\ref{eq:EOSisentropic}) with $K=100$ and $\Gamma=2$.

The results of this comparison are summarised for model U11 in
Fig.~\ref{fig:U11adia} and indicate that the non-isentropic changes
are indeed very small and that these do not produce any significant
variations on the development of the instability and on the growth of
the $m$=2 mode. Larger differences are seen in the growth of the 
$m$=1 mode, but also these are very small and do not produce a qualitative
change. Quantitative assessment of the changes produced by a different
EOS are reported in Table~\ref{table:results}, but these are, overall,
comparable with the error bar for the determination of $\tau_{_{\rm
    B}}$ and $f_{_{\rm B}}$. Finally, all the considerations made
here for model U11 apply also to models U13 and U3, with model U3
being slightly more sensitive to the change in EOS (\textit{cf.}
Table~\ref{table:results}). These results indicate therefore that the
effects of shock heating are likely to be unimportant at least for the
development and evolution of the bar in isolated and old neutron
stars.

\begin{figure}
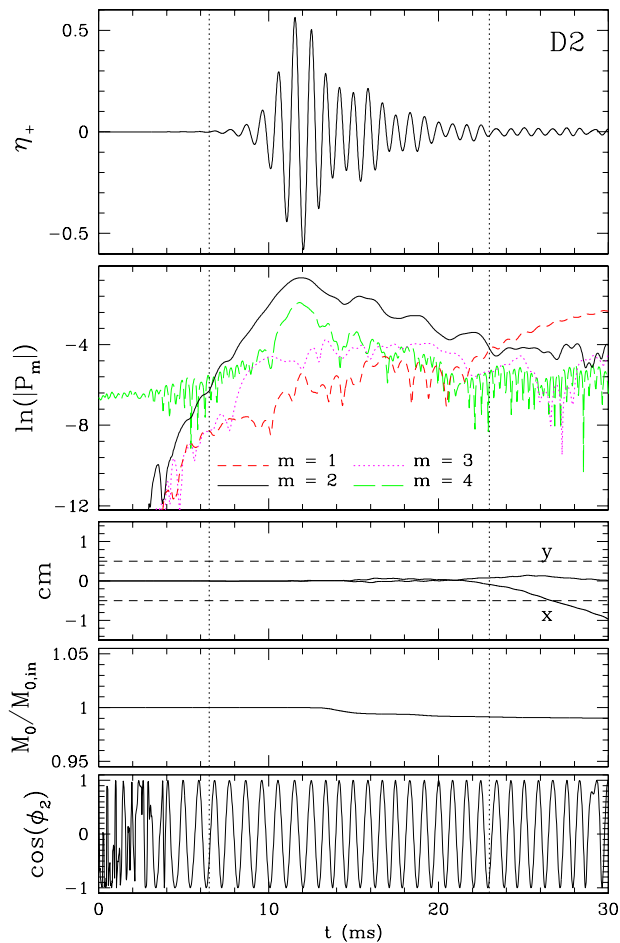

\begin{center}
\mbox{\includegraphics[width=8.6truecm]{\imgname{eta20}}}\\
\mbox{\includegraphics[width=8.6truecm]{\imgname{md20}}}\\
\mbox{\includegraphics[width=8.6truecm]{\imgname{cm20}}} \\
\mbox{\includegraphics[width=8.6truecm]{\imgname{rm20}}}\\
\mbox{\includegraphics[width=8.6truecm]{\imgname{fs20}}}\\
\end{center}
\vspace{-0.7cm}
\caption{The same as in Fig.~\ref{fig:U3pisym} but for model D2. In
  addition, the bottom panel shows the evolution of the phase for the
  $m=$2 mode.}
\label{fig:D2}
\end{figure}

\subsection{The role of grid spacing and size}
\label{sec:effectRES}

We finally report on the influence of the grid spacing and of the
grid size on the development of the instability. To assess this we
have performed several simulations of the intermediate model {U11}
differing either in grid resolution or in the location of the outer
boundaries. In particular, we have considered grid resolutions $\Delta
x/M_\odot=0.375$, $\Delta x/M_\odot=0.5$ and $\Delta x/M_\odot=0.625$ 
and found the code to be second-order convergent, with the coarsest 
resolution being just on the limit of the convergence regime 
(the results with $\Delta x/M_\odot=0.75$ are in fact not convergent 
at the expected rate).

As summarized in Table~\ref{table:results} we have found that the
computed values of the instability parameters $\tau_{_{\rm B}}$ and
$f_{_{\rm B}}$ do not vary significantly across the range of
resolutions considered, with differences that are at most of about
$3\%$. The same Table also contains information on the results
obtained when comparing simulations performed with $\Delta
x/M_\odot=0.625$ but with a larger computational domain, namely going
from a computational box, at this resolution, with extents $[157\times
  157\times 56]$ to one with extents $[211\times 211\times 80]$.  Also
in this case the changes in the dynamics are very small (see also the
right panel of Fig.~\ref{fig:U11_cf}) and essentially amount to a
smaller loss of mass and angular momentum as some of the the matter in the spiral
arms is thrown out of the computational grid (see also discussion in
Sec.~\ref{sec:persistence}).


\begin{table*}
\begin{tabular}{|c|c|cc|ccc|ccc|cc|c|c|}
\hline\hline 
Model   &$\Delta x/M_\odot$  &$\delta_1$ & $\delta_2$ 
        & EOS & $\pi$-sym   
        & grid size &$\Delta t$&$t_1$&$t_2$ 
        &${\tau}_{_{\rm B}}$&$f_{_{\rm B}}$ 
        & $\eta$ &$\tau_{_{\rm D}}$ \\
     &       &       &       &           &     &             & (ms)  & (ms)& (ms)&  (ms) & (Hz) &(max) & (ms) \\
\hline\hline 
U3   & 0.500 & $0.0$ & $0.0$ &polytropic      & no  & medium &  0.76  & 21  & 26  & 2.79  &  552 & 0.48 & 13.8\\
U3   & 0.500 & $0.0$ & $0.0$ & ideal fluid    & no  & medium &   0    & 21  & 26  & 2.69  &  547 & 0.47 & 12.9 \\
U3   & 0.500 & $0.0$ & $0.01$& ideal fluid    & no  & medium & -15.94 & 21  & 26  & 2.42  &  548 & 0.53 & 13.0 \\
U3   & 0.625 & $0.0$ & $0.0$ & ideal fluid    & yes & medium &  -1.00 & 21  & 26  & 2.82  &  543 & 0.43 & 24.4 \\
U3   & 0.625 & $0.0$ & $0.01$& ideal fluid    & yes & medium & -16.28 & 21  & 26  & 2.52  &  547 & 0.54 & 24.5 \\
\hline  
U11  & 0.500 & $0.0$ & $0.0$ & polytropic     & no  & medium &  5.35 & 11   & 14  & 1.12  &  497 & 0.78 &  8.6 \\
U11  & 0.500 & $0.0$ & $0.0$ & ideal fluid    & no  & medium &   0    & 11  & 14  & 1.15  &  494 & 0.78 &  9.4 \\
U11  & 0.500 & $0.0$ & $0.01$& ideal fluid    & no  & medium &  -8.55 & 11  & 14  & 1.11  &  494 & 0.79 &  9.9 \\
U11  & 0.375 & $0.0$ & $0.0$ & ideal fluid    & no  & medium &   1.64 & 11  & 14  & 1.11  &  492 & 0.79 & 10.5 \\
U11  & 0.625 & $0.0$ & $0.0$ & ideal fluid    & no  & medium &   1.79 & 11  & 14  & 1.15  &  492 & 0.78 &  9.0 \\
U11  & 0.625 & $0.0$ & $0.0$ & ideal fluid    & no  & large  &   2.54 & 11  & 14  & 1.15  &  492 & 0.78 &  9.8 \\
U11  & 0.625 & $0.0$ & $0.0$ & ideal fluid    & yes & medium &   1.39 & 11  & 14  & 1.12  &  494 & 0.77 & 13.8 \\
U11  & 0.750 & $0.0$ & $0.0$ & ideal fluid    & no  & medium &   3.79 & 11  & 14  & 1.17  &  493 & 0.76 & 10.8 \\
U11  & 0.625 & $0.05$& $0.0$ & ideal fluid    & no  & medium &   6.20 & 11  & 14  & 1.14  &  495 & 0.78 &  6.6 \\
\hline 
U13  & 0.500 & $0.0$ & $0.0$ & polytropic     & no  & medium &   1.69 & 10  & 13  & 0.94  &  457 & 0.86 & 5.7 \\
U13  & 0.500 & $0.0$ & $0.0$ & ideal fluid    & no  & medium &   0    & 10  & 13  & 0.95  &  454 & 0.85 & 6.2 \\
U13  & 0.500 & $0.0$ & $0.01$& ideal fluid    & no  & medium &  -8.55 & 10  & 13  & 0.93  &  454 & 0.86 & 6.3 \\
U13  & 0.625 & $0.0$ & $0.0$ & ideal fluid    & yes & medium &  -0.16 & 10  & 13  & 0.96  &  453 & 0.86 & 6.5 \\
U13  & 0.625 & $0.0$ & $0.01$& ideal fluid    & yes & medium &  -8.71 & 10  & 13  & 0.94  &  454 & 0.86 & 6.2 \\
\hline\hline 
D2   & 0.500 & $0.0$ & $0.0$ & ideal fluid    & no  & medium &   0    &  9  & 10.5& 0.90  & 1053 & 0.59 & -- \\
D2   & 0.500 & $0.0$ & $0.01$& ideal fluid    & no  & medium &  -6.54 &  9  & 10.5& 0.78  & 1052 & 0.67 & -- \\
D2   & 0.500 & $0.0$ & $0.04$& ideal fluid    & no  & medium &  -7.57 &  9  & 10.5& 0.77  & 1056 & 0.67 & -- \\
\hline 
D3   & 0.500 & $0.0$ & $0.0$ & ideal fluid    & no  & medium &   0    &  9  & 12  & 1.54  & 1086 & 0.38 & -- \\
D7   & 0.500 & $0.0$ & $0.0$ & ideal fluid    & no  & medium &   0    & 11  & 14  & 1.74  &  821 & 0.48 & -- \\
\hline\hline 
\end{tabular}

\caption{Main properties of the initial part of the instability for
  the stellar models used in the simulations. Starting from the left
  the different columns report: the grid spacing $\Delta x/M_{\odot}$,
  the amplitudes of the initial perturbations in the $m$=1 and
  $m$=2 modes $\delta_{1,2}$, the EOS, the symmetry and the grid size
  used, the time shift $\Delta t$ [\textit{cf.}
    eq. \eqref{eq:defSHIFT}], the times $t_1$ and $t_2$ between which
  the growth-times $\tau_{_{\rm B}}$ and the frequencies $f_{_{\rm
      B}}$ are computed, the maximum value of the distortion parameter
  $\eta$, and the duration of the bar deformation $\tau_{_{\rm D}}$.}
\label{table:results}
\end{table*}

\subsection{Comparison with previous studies}
\label{sec:shibata2000}

To conclude this Section describing the dynamics of the instability, we
comment on the important validation of the accuracy of our simulations
that comes from a comparison with results previously published in the
literature. We have focused, in particular, on the fully general-relativistic 
simulations published in ref.~\cite{Shibata00b} and
repeated those relative to the models {D2}, {D3} and {D7} discussed
there. These stars have instability parameters $\beta$ rather close to
the critical one, but are also more massive, with gravitational masses
between 2 and 2.6 $M_{\odot}$, and have larger compactnesses
(\textit{cf.} Table~\ref{table:models}). The development of the
instability for one of these models (D2) is summarised in
Fig.~\ref{fig:D2} and the computed distortion parameter is
qualitatively very similar to the one presented in
ref.~\cite{Shibata00b} and shows that in this compact star the bar 
[\textit{i.e.} stage \textit{(b)} of the classification made in 
Sec.~\ref{sec:unstable}] is very rapidly attenuated as a result of the
development of spiral arms, which are also responsible for a small
loss of mass.

The frequencies and the growth times found for these models are also
in good agreement with those reported in ref.~\cite{Shibata00b}, but
are not identical; differences are of about 10\% (see
Table~\ref{table:results} for a close comparison). While there are
several differences in the numerical codes used, it should be noted
that the simulations reported in ref.~\cite{Shibata00b} made use of a
substantial perturbation in the $m$=2 mode, with an equivalent
$\delta_2=0.3$. Although we were not able to reproduce exactly the
dynamics of these models (no convergent solution of the constraint
equations was found once such a large perturbation was introduced), we
recall that large perturbations for models near the threshold do
induce a change in the growth rates and effectively reduce the growth
times (see discussion in Sec.~\ref{sec:effectPERT}).

We believe therefore that the use of a smaller or zero perturbation is
the largest source of the difference with the corresponding
simulations in ref.~\cite{Shibata00b}, which we can nevertheless
reproduce to very good precision.

\section{Determination of the threshold}
\label{sec:threshold}

An important consequence of the high accuracy of our simulations it that
it has allowed for a rather precise determination of the stability
threshold for the sequence of models considered here. Of course, the
determination of the threshold to the third significant figure has little
but academic interest, as it is expected not to be universal, but rather to
depend (although weakly) on properties such as the degree of differential
rotation, the compactness, the EOS, etc.. Nevertheless, this is a useful
exercise for at least two different reasons. Firstly, it helps in
characterizing the dynamics of slightly supercritical models and,
secondly, it allows for a direct comparison with perturbative studies,
highlighting when the latter cease to be accurate and how they can be
improved.

In what follows we discuss two different methods we have used to this
aim. The first one simply selects the initial model with the lowest
value of $\beta$ for which an exponential growth in the distortion
parameter is seen. The second one, on the other hand, uses the classical
Newtonian stability analysis of Maclaurin spheroids for incompressible
self-gravitating fluids in equilibrium~\cite{Chandrasekhar69} to
extrapolate the position of the threshold also in a fully
general-relativistic regime. Interestingly, the results of the two
approaches agree to high precision.

\subsection{First method: dynamical evaluation}
\label{subsec:threshold1}

The approach followed here is straightforward and consists in
performing a number of simulations for models with decreasing values
of the instability parameter $\beta$ and in determining the critical value
$\beta_c$ as the smallest one for which an exponential growth of the
distortion parameter $\eta$ is observed. More specifically, we have
performed a set of 10-ms simulations of models characterized by values
of $\beta$ between $\beta=0.24$ and $\beta=0.255$, interval in which
the instability was reported to develop~\cite{Shibata00b, Saijo2000,
  Saijo:2001}. In order to remove a possible contamination by the
Cartesian discretization through the power in the $m$=4 mode, all 
the models had a very small $m$=2-mode perturbation with $\delta_2=0.04$,
making this the largest mode power initially. Furthermore, since this
kind of initial $m$=2-mode perturbation always generates, at least
temporarily, a growth of the distortion, we classified as unstable
those models for which an exponentially growing bar deformation was
observed for the whole simulated 10 ms.

\begin{figure}
  \begin{center}
  \vspace{-0.4cm}
  \includegraphics[width=8.5truecm]{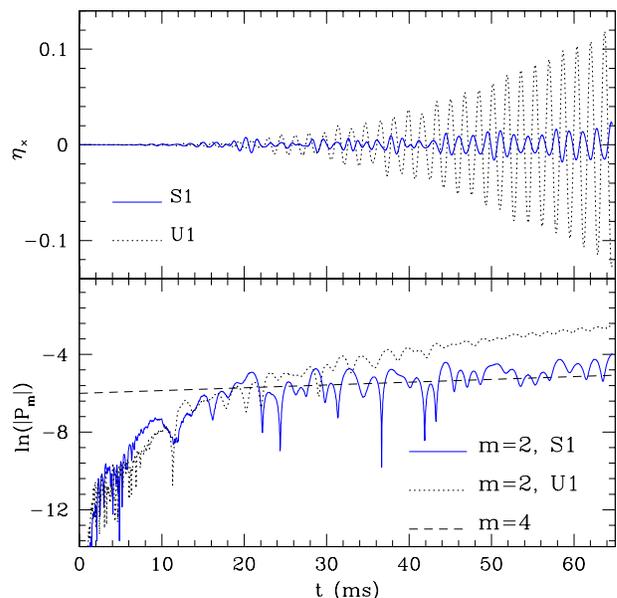}
  \vspace{-1.0cm}
  \end{center}
    \caption{Dynamics of the instability near the threshold. The two
      panels summarise two long-term evolutions for model {S1} (continuous
      line) and {U1} (dotted line) and show the distortion parameter
      and the power in the $m$=2 and $m$=4 modes.}
\label{fig:DeterminationInstabilityNoPert}
\end{figure}

As a result of this set of simulations and of additional refinements
of the bracketing interval for the instability, we have concluded that
the threshold had to be found between models U2 and S2, although it
was not yet obvious whether any or all of these models were
unstable. A more precise determination of the nature of these models
has therefore required much longer simulations to be performed and
that no initial perturbation was introduced. A summary of these
long-term simulations is reported in
Fig.~\ref{fig:DeterminationInstabilityNoPert}, which shows the
evolution for models U1 and S1. The upper panel, in particular, shows
the amplitude of the distortion parameter $\eta_{\times}$ for model S1
(continuous line) and U1 (dotted line), respectively, while the lower
panel shows the evolution of the power in the $m$=2 mode; indicated
with a dashed line is the average value of the $m$=4-mode power for
model S1, which does not show an appreciable growth. 

While it is evident that model U1 develops an instability over the
timescale of the simulation, S1 does not, implying that the threshold
for the onset of the dynamical bar-mode instability for the sequence
under investigation is $\beta_c\simeq 0.255$. We are aware that it may be
argued that model S1 is also unstable but with a much smaller growth
rate; assessing in practice this would require simulations which are
prohibitive with the present computational facilities. However,
confidence that the result obtained is correct is also provided by
critical-fit analysis, which will be discussed in the following
Section.

\begin{table}
\begin{tabular}{|l|c|c|cc|c|cc|}
\hline
Model& $\beta$&$\delta_2$&$t_1$&$t_2$ & \multicolumn{1}{|c|}{$\eta$}& $\tau_{_{\rm B}}$
                                      &\multicolumn{1}{c|}{$f_{_{\rm B}}$} \\
     &        &        &(ms)&(ms)&(max)    &  (ms) & \multicolumn{1}{c|}{(Hz)}  \\
\hline 
S6   & 0.240  & $0.04$ &  3  &  9   & 0.02$  $& --   & $740_{-64  }^{+145}$\\
S5   & 0.245  & $0.04$ &  3  &  9   & 0.02$  $& --   & $705_{-64  }^{+123}$\\
S4   & 0.250  & $0.04$ &  3  &  9   & 0.03$  $& --   & $656_{-65  }^{+71 }$\\
S3   & 0.252  & $0.04$ &  3  &  9   & 0.04$  $& --   & $611_{-30  }^{+43 }$\\
S2   & 0.253  & $0.04$ &  3  &  9   & 0.05$  $& --   & $588_{-12  }^{+25 }$\\
S1   & 0.254  & $0.04$ &  3  &  9   & $*$     & 9.71 & $578_{-21  }^{+15 }$\\
U1   & 0.255  & $0.04$ &  3  &  9   & $*$     & 5.26 & $567_{-4   }^{+6  }$\\
\hline 
S1   & 0.254  & $0.0$  & 45  & 63   & 0.02$  $& --   & $599_{-209 }^{+209}$\\
U1   & 0.255  & $0.0$  & 45  & 63   & $*$     & 22.1 & $588_{-96  }^{+56 }$\\
\hline
\end{tabular}
\caption{The Table reports for some models near the instability threshold: the ratio $\beta$ of the
   rotational kinetic energy to the gravitational binding energy; the initial $m$=2-mode
   perturbation $\delta_2$; the bounds $t_1$ and $t_2$ of the considered time interval; the maximum
   distortion $\eta$; the growth rate $\tau_{_{\rm B}}$ and the frequency $f_{_{\rm B}}$ of the
   bar-mode during the initial part of the instability. For these models, $\Delta x/M_\odot=0.5$ and
   $\pi$-symmetry is not used. The growth rate $\tau_{_{\rm B}}$ and the frequency $f_{_{\rm B}}$ of
   the bar-mode are obtained making a least-square fit of $\eta_+(t)$ (eq.\ref{eq:etafit}) between
   $t_1$ and $t_2$. The $*$ indicates that $\eta$ did not reach a
   maximum before the simulation was stopped.}
\label{table:criticalpert}
\end{table}

\subsection{Second method: critical fit}
\label{subsec:threshold2}

The classical Newtonian study of the bar-mode instability is based on
the stability analysis of Maclaurin spheroids for an incompressible
and self-gravitating fluid in equilibrium~\cite{Chandrasekhar69}. We
recall that by considering the linearized form of the second-order
virial equation, which governs the small oscillations around the
equilibrium configuration, it is possible to show that the toroidal
$m$=2 perturbations have complex eigenvalues given by (we here use
Chandrasekhar's notation of ref.~\cite{Chandrasekhar69})
\begin{equation}
\sigma=\Omega(e) \pm \sqrt{4B_{11}(e)-\Omega^2(e)}
\quad ,
\label{eq:CHANDRA}
\end{equation}
where $e$ is the eccentricity of the Maclaurin spheroids relative to
an incompressible Newtonian star ($e=0$ for a spherical
star) and 
\begin{eqnarray}
B_{11}   \!&=&\!\frac{3\,e - 5\,e^3 + 2\,e^5 + {\sqrt{1 - e^2}}\,
                  \left(4\,e^2 -3\right) \arcsin (e)
                 }{4\,e^5}\ ,
\nonumber\\
\Omega^2\!&=&\!\frac{6\,\left(e^2 -1\right) }{e^2} + 
               \frac{2\,\left( 3 - 2\,e^2 \right) 
                   \,{\sqrt{1 - e^2}}\,\arcsin (e)
                  }{e^3}\ .
\nonumber\\
\end{eqnarray}
As customary in perturbative studies, the real part of the
eigenfrequency $\sigma$ represents the characteristic frequency of the
small perturbation, while its imaginary part measures the exponential
growth time of the $m$=2 mode and is nonzero if
$4B_{11}(e)-\Omega^2(e)<0$, with the threshold for the instability
being marked by the value of the eccentricity for which
$4B_{11}(e)-\Omega^2(e)=0$. In the case of an incompressible Newtonian
star, the eccentricity $e$ and the instability parameter $\beta$ are
related as
\begin{equation}
\beta(e)=-1+\frac{3}{2e^2}-\frac{3\sqrt{1-e^2}}{2e \arcsin(e)}\ .
\end{equation}
Equation~(\ref{eq:CHANDRA}) is most conveniently rewritten in terms of
the instability parameter $\beta$ as 
\begin{equation}
\sigma =\Omega(\beta) \pm \frac{{\mathrm i}}{\tau(\beta)}
\quad ,
\label{eq:CHANDRAbis}
\end{equation}
where 
\begin{equation}
1/\tau^2 \equiv -4B_{11}(\beta)+\Omega^2(\beta)\ .
\end{equation}

\begin{table}[t]
\begin{center}
\begin{tabular}{|l|c|rr|r|ll|}
\hline
Model& $\beta$&$t_1$&$t_2$ & \multicolumn{1}{|c}{$\eta$}
                           & \multicolumn{1}{|c}{${\tau}_{_{\rm B}}$}
                           & \multicolumn{1}{c|}{$f_{_{\rm B}}$} \\
     &        & (ms) & (ms) &(max)   
                           & \multicolumn{1}{|c}{(ms)}
                           & \multicolumn{1}{c|}{(Hz)} \\
\hline 
U2   & 0.2581 & 16.9 & 22.4 & 0.3734 & $3.438_{-1.26}^{+2.39}$&  $552_{ -30}^{ +72}$   \\
U3   & 0.2595 & 19.9 & 24.2 & 0.4241 & $2.678_{-1.23}^{+1.33}$&  $544_{ -39}^{ +86}$   \\
U4   & 0.2621 & 15.3 & 18.3 & 0.5496 & $1.854_{-0.17}^{+0.13}$&  $540_{ -12}^{ +14}$   \\
U5   & 0.2631 & 16.2 & 19.0 & 0.5788 & $1.748_{-0.10}^{+0.12}$&  $538_{ -13}^{ +6 }$   \\
U6   & 0.2651 & 14.5 & 17.1 & 0.6305 & $1.574_{-0.04}^{+0.05}$&  $528_{ -3 }^{ +6 }$    \\
U7   & 0.2671 & 14.2 & 16.4 & 0.6694 & $1.408_{-0.06}^{+0.06}$&  $522_{ -15}^{ +8 }$   \\
U8   & 0.2686 & 12.2 & 14.3 & 0.7027 & $1.319_{-0.03}^{+0.06}$&  $518_{ -6 }^{ +3 }$    \\
U9   & 0.2701 & 13.2 & 15.2 & 0.7223 & $1.269_{-0.04}^{+0.05}$&  $512_{ -1 }^{ +3 }$    \\
U10  & 0.2721 & 13.7 & 15.6 & 0.7482 & $1.184_{-0.03}^{+0.05}$&  $503_{ -7 }^{ +18}$    \\
U11  & 0.2743 & 12.9 & 14.7 & 0.7749 & $1.116_{-0.01}^{+0.04}$&  $493_{ -11}^{ +4 }$   \\
U12  & 0.2761 & 12.0 & 13.7 & 0.7999 & $1.066_{-0.03}^{+0.03}$&  $486_{ -6 }^{ +5 }$    \\
U13  & 0.2812 & 11.2 & 12.7 & 0.8551 & $0.952_{-0.02}^{+0.02}$&  $453_{ -7 }^{ +5 }$    \\
\hline
\end{tabular}
\caption{Same quantities as in Table \ref{table:criticalpert}, but referring to the models evolved
  with a grid size of $\Delta x/M_\odot=0.625$, with $\pi$-symmetry and no perturbation.  The growth
  rate $\tau_{_{\rm B}}$ and the frequency $f_{_{\rm B}}$ of the bar-mode are obtained making a
  least-square fit of $\eta_+(t)$ (eq.\ref{eq:etafit}) between $t_1$ and $t_2$. The interval
  $[t_1,t_2]$ is here determined, differently from the previous tables, as the one in which
  $\eta(t)$ is between 5\% and the 25\% of its first maximum. These models were used to determine
  the threshold.}
\label{table:critical}
\end{center}
\end{table}

With this in mind it is natural to ask how accurate is the Newtonian
description of an incompressible Maclaurin spheroid in describing the
nonlinear dynamics of a relativistic differentially rotating star. Of
course also in full General Relativity the transition between stable
and dynamical unstable bar-modes will be described by a change from real
to imaginary of the $m$=2-mode eigenfrequency and it is therefore
reasonable to expect that the frequencies and the growth times depend
on $\beta$ as
\begin{eqnarray}
\frac{\Omega(\beta)}{2\pi} 
       &\approx& f_c+f^{(1)}_c(\beta-\beta_c)+ f^{(2)}_c(\beta-\beta_c)^2  
\ ,\label{eq:fit Omega}\\
\frac{1}{\tau^2} 
       &\approx& \frac{1}{k^2} (\beta-\beta_c)
\quad . \label{eq:fit beta}
\end{eqnarray}
Using this \textit{ansatz}, we have performed a series of simulations
for models U2--U13 using a $\pi$-symmetry and a grid resolution of
$\Delta x/M_\odot=0.625$, through which we have computed the values for
$f_{_{\rm B}}$ and $\tau_{_{\rm B}}$ by means of a nonlinear
least-square fit to the trial form of eq.~(\ref{eq:etafit}). Making
use of these results, which are collected in
Table~\ref{table:critical}, we have then again used a least-square fit
of the values $f_{_{\rm B}}$ and $\tau_{_{\rm B}}$ to the expected
$\beta$ dependence of eqs.~(\ref{eq:fit Omega}) and (\ref{eq:fit
  beta}) to obtain the unknown coefficients $f_c, f^{(1)}_c,
f^{(2)}_c$, $k$ and $\beta_c$:
\begin{equation}
\begin{aligned}
 f_c        &=   554\, \mbox{Hz}\ , & 
 f^{(1)}_c\! &= -1668\, \mbox{Hz}\ , &
 f^{(2)}_c\! &= -85635\, \mbox{Hz}\ ,  \\
\nonumber \\
 \beta_c  &= 0.2554\ , &
 k        &= 0.153\ \mbox{ms}\ .
\end{aligned}
\end{equation}

\begin{figure*}[tbf]
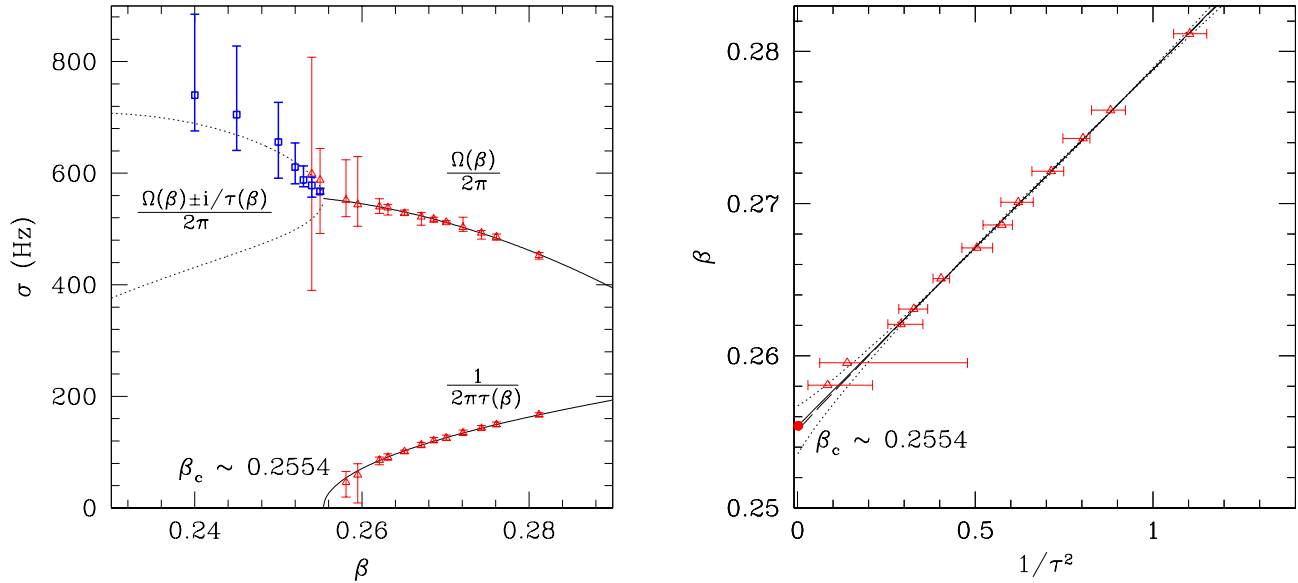

\begin{center}
\includegraphics[width=8.5truecm]{\imgname{freqfit}}
\hskip 0.5cm
\includegraphics[width=8.5truecm]{\imgname{critical_fit}}
\end{center}
\vglue-0.9cm
\caption{{\it Left panel:} critical diagram as constructed with the
  frequencies and growth times relative to the unperturbed models of
  Table~\ref{table:critical} (triangles) and to the perturbed models
  of Table~\ref{table:criticalpert} (squares). The continuous lines
  represent the two fitted curves for $\Omega(\beta)$ and
  $\tau(\beta)$, while the dotted line the corresponding
  extrapolations below the threshold. {\it Right panel:} same data as
  in the left panel but magnified around the critical threshold and
  expressed in terms of $1/\tau^2$ to highlight the very good fit.  }
\label{fig:freqfit} \label{fig:critical_fit}
\end{figure*}

In the left panel of Fig.~\ref{fig:critical_fit} we summarize the
results of this fit for the models U2--U13 and show the corresponding
error bars. In particular, we denote with triangles the unperturbed
models reported in Table \ref{table:critical}, while with squares the
perturbed models in Table \ref{table:criticalpert}. In addition, the
continuous lines represent the two fitted curves for $\Omega(\beta)$ and
$\tau(\beta)$, while the dotted line refers to the corresponding
extrapolations below the threshold. We note that the error bars in
Fig.~\ref{fig:critical_fit} are computed in different ways for the growth rates and the
frequencies. In the first case they are computed as the difference
between the minimum and maximum values of $d\log(\eta(t))/dt$ in the
time intervals between $t_1$ and $t_2$ reported in
Table~\ref{table:critical}. In the second case, instead, the error
bars are determined using the minimum and maximum values over the time
intervals between $t_1$ and $t_2$ reported in
Tables~\ref{table:criticalpert} and \ref{table:critical} of the
pattern speeds extracted from the collective phase $\phi_2(t)$ of
eq.~(\ref{eq:phimodes}).

A number of comments are worth making. Firstly, it is clear that
the Newtonian description of the instability in terms of
incompressible Maclaurin spheroids is surprisingly accurate also in
full General Relativity and for differentially rotating stars. It is
especially so for models which are far from the instability threshold,
for which the error bars are very small and below 5\%. Secondly, as
the models approach the critical threshold from above (\textit{i.e,}
unstable models), the growth times become increasingly large and the
numerical errors increasingly more important for the resolutions we
could use. Yet, even in this regime the perturbative predictions are
accurate to better than 15\%. Finally, as the models approach the
critical threshold from below (\textit{i.e,} stable models), the frequencies of the
oscillations triggered by the use of an initial perturbation are similar
to the spin frequencies, making the determination of the
eigenfrequencies increasingly difficult and hence inaccurate
(\textit{cf.} the large error bars for models denoted by squares in
the left panel of Fig.~\ref{fig:freqfit}).

Finally we note that the independent determination of the threshold
for the instability provided by the fit to eq.~(\ref{eq:fit beta})
(\textit{i.e.} $\beta_c = 0.2554$) is in surprisingly good agreement
with the one obtained in the previous Section, \textit{i.e.} $\beta_c
\simeq 0.255$, confirming the accuracy of the dynamical determination
and suggesting that model S1 is indeed stable. The very good fit of
the data is shown in the right panel of Fig.~\ref{fig:critical_fit},
 which reports the same data as in the left panel but magnified
around the critical threshold and shown in terms of $1/\tau^2$ to
highlight the functional dependence as expressed by eq.~(\ref{eq:fit
  beta}).

\section{Gravitational-Wave emission and detection of the 
            unstable models}
\label{sec:GW}

One of the goals of this study is to assess whether a dynamical bar-mode
instability triggered in a neutron-star model could be a good source of
gravitational radiation for the detectors presently collecting data
(LIGO, GEO), in the final stages of construction (Virgo) or in the
planning phase (Dual)~\cite{Bonaldi:Dual2003}. The use of uniform grids
in these calculations has prevented us from placing the outer boundaries
at distances sufficiently large to allow a gauge-invariant extraction of
the gravitational waves~\cite{Abrahams97a, Rupright98,
Rezzolla99a}. However, a reasonable measure of the amplitude of the
expected gravitational radiation and of the consequent SNR is still
possible by making use of the standard quadrupole
formula~\cite{Misner73}. A number of comparative tests have been carried
out among the various ways in which the gravitational-wave amplitudes can
be estimated (see, for instance, refs.~\cite{Tanaka93,
ShibataSekiguchi2004,nagar:05}) and we expect the error in this case to
be ${\cal O}(M/R)$~\cite{Shibata03d} and thus of a few percent only.

In this approximation, the observed waveform and amplitude for the two
polarizations measured by an observer situated at infinity along the
$z$-axis are
\begin{equation}
\label{eq:GWamplitude}
\begin{split}
h_{+}     &=\dfrac{\ddot{I}^{xx}(t')-\ddot{I}^{yy}(t')}{r} \ ,\\
h_{\times} &=2\; \dfrac{\ddot{I}^{xy}(t')}{r} \ ,
\end{split}
\end{equation}
where $t'=t-r$ is the retarded time. The total gravitational wave
luminosity is then given by
\begin{equation}
\label{eq:GWluminosity}
\dot{E} =\dfrac{1}{5} \big\langle 
      \dddot{I}^{\,ij} \dddot{I}_{ij} \big\rangle \ ,
\end{equation}
while the angular momentum loss is
\begin{equation}
\label{eq:GWAngularMomentum}
\dot{J_i} =\dfrac{2}{5} \big\langle 
      \epsilon_{ijk} \ddot{I}^{jl} \dddot{I}_{lk}
\big\rangle \ .
\end{equation}

We note that the definition of the quadrupole tensor is not unique as
$\rho$ and $D$ coincide in a Newtonian approximation and either of them
could be used. Following ref.~\cite{Shibata00b}, we adopt the definition
expressed in eq.~(\ref{eq:defQuadrupole}), as this allows us to exploit
the conservation equation for $D$ to perform analytically the first time
derivative of the quadrupole tensor and obtain that
\begin{equation}
\label{eq:DQuadrupole}
\dot{I}^{ij} = \int d^{3}\!x \; D\; 
    \left[ x^{i} (\alpha v^{j} + \beta^{j})
             +x^{j} (\alpha v^{i} + \beta^{i})
    \right] \ ,
\end{equation}
from which we compute the second $\ddot{I}^{ij}$ and third
$\dddot{I}^{\,ij}$ time-derivatives using first-order finite
differencing. This approach not only is simpler, but it is indeed the
only possible one, since our second-order--accurate scheme would not
allow for an accurate direct calculation of $\dddot{I}^{\,ij}$ (see
discussion in Appendix A of ref.~\cite{Rezzolla99a}).

\begin{figure}[t]
\begin{center}
\includegraphics[width=7.75truecm]{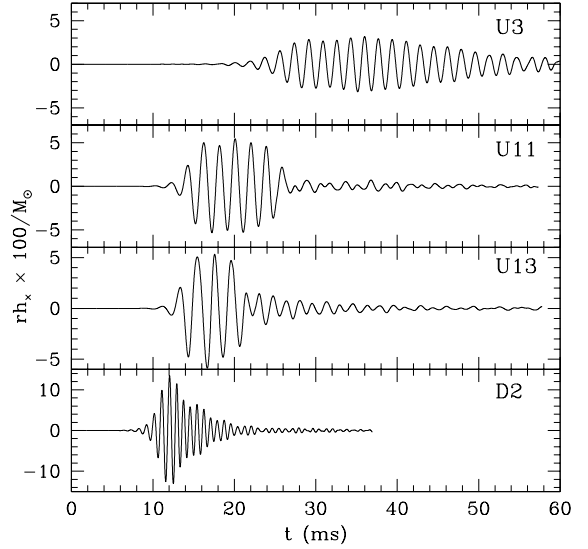}
\end{center}
\vspace{-.7cm}
\caption{Gravitational-wave signals along the $z$-axis for the
  ``cross'' polarization as computed in the Newtonian quadrupole
  approximation for models {U3}, {U11}, {U13} and {D2}.}
\label{fig:GW}
\end{figure}

In Fig.~\ref{fig:GW} the we show the waveforms for the ``cross''
polarization computed in this way, as measured along the $z$-axis for
models {U3}, {U11}, {U13} and {D2} (We recall that the ``cross'' and
``plus'' waveforms for a stationary and rotating bar differ only by a
phase, since the emission is circularly polarized). Clearly, the
gravitational-wave signal mimics (modulo a factor of 2 in the
frequency) the dynamical behaviour of the bar deformation, with
signals that are longer for models near the stability threshold ({\it
  i.e.} model U3), and with amplitudes which increase for the more
compact model D2.

\begin{figure*}[th]
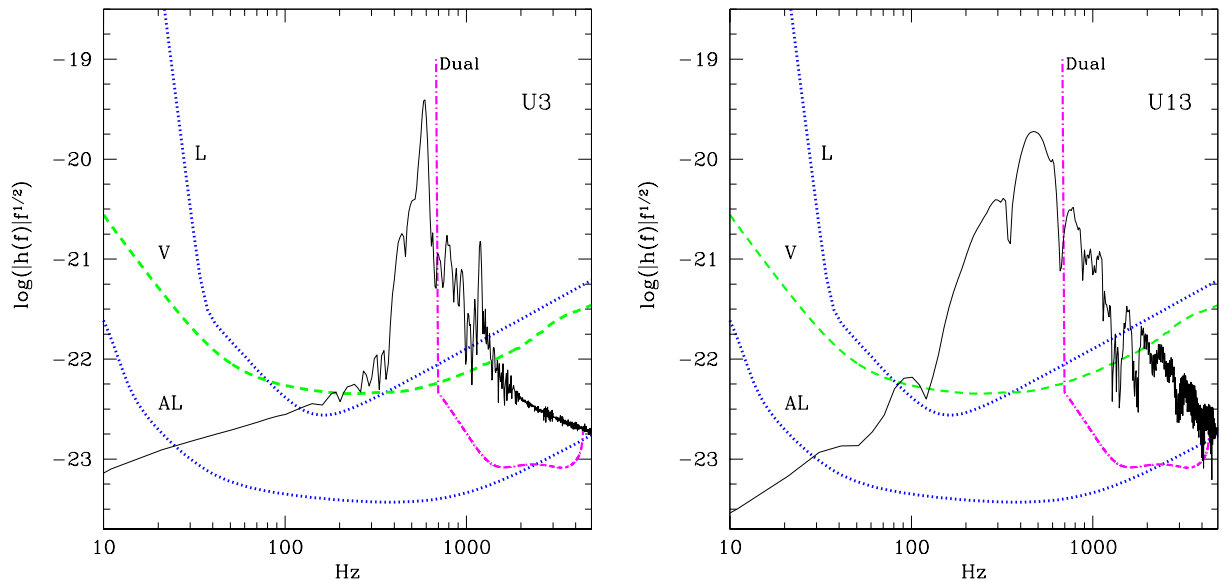

\begin{center}
\includegraphics[width=8.25truecm,height=8.25truecm]{\imgname{gw_spectrumU3}}
\includegraphics[width=8.25truecm,height=8.25truecm]{\imgname{gw_spectrumU13}}
\end{center}
\vspace{-.6cm}
\caption{Comparison between $|\tilde{h}(f)| f^{1/2}$ for models U3 and
  U13 at 10kpc and the square-root of the power spectrum of the noise of
  Virgo (dashed line), LIGO (dotted line), Advanced LIGO (dotted-line)
  and the planned resonant detector Dual (dot-dashed line). Note the
  significant difference in the power spectrum of the two signals, with
  the one relative to model U3 having a larger and narrower peak at about
  600~Hz, produced by the more persistent bar deformation.}
\label{fig:SNRLIGO}
\end{figure*}
\begin{table}[t]
\begin{tabular}{|c|cc|c|cccc|}
\hline
Model& $\Delta M/M$ & $\Delta J/J$ & $h_{\rm rss}$  & $SNR$  & $SNR$ & $SNR$ & $SNR$ \\  
     & $\scriptstyle (10^{-5})$  
     & $\scriptstyle (10^{-3})$    
     & $\scriptstyle (10^{-20})$   & V  & L& AL & Dual \\
\hline                      
U3   & $7.75$  & $3.82$ & $1.39$  & 117 &  82.5 & 1590 & 19 \\
U11  & $9.48$  & $4.84$ & $1.74$  & 155 & 119   & 2040 & 23 \\
U13  & $6.23$  & $3.18$ & $1.51$  & 138 & 114   & 1780 & 30 \\
D2   & $52.6$  & $10.2$ & $2.50$  & 136 &  79.8 & 2270 & 892 \\
\hline                      
\end{tabular}
\caption{List of the representative gravitational-wave quantities
  computed in the Newtonian quadrupole approximation for models
  U3, U11, U13 and D2. From the left the different columns report:
  the fractional amounts of the energy and angular momentum carried by
  the gravitational radiation ($\Delta M/M$ and $\Delta J/J$,
  respectively), the root-sum-square of $h_+$ for a source at 10 kpc,
  and the SNRs for Virgo, LIGO, Advanced LIGO and Dual.}
\label{table:GWamplitude}
\end{table}

A convenient measure of the strength of the signal can be given in
terms of the root-sum-square amplitude of, say, the plus polarization
\begin{equation}
h_{\rm rss}=\left[\int_{-\infty}^{+\infty}\!\!\!\! dt \, h_{+}^2(t)\right]^{1/2}
\ ,
\label{eq:hrss}
\end{equation}
since $h_{\rm rss}$ has the same units as the strain-noise amplitude
of the detectors and it is therefore possible to obtain a rough
estimate of the SNR simply dividing it by the strain-noise amplitude
at the frequency where the signal is the strongest. In
Table~\ref{table:GWamplitude} we report the values of $h_{\rm rss}$
for a signal coming from a source at 10 kpc, together with the SNR
computed assuming the optimal use of match-filtering techniques and
given by
\begin{equation}
\frac{S}{N}=2 \sqrt{\int_{0}^{+\infty}\!\!d \log f \; 
  \frac{|\tilde{h}_{+}(f)|^2 f}{S_n(f)}} \quad ,
\label{eq:snr2}
\end{equation}
where $\tilde{h}_{+}(f)$ is the Fourier transform of $h_+(t)$ and
$S_n(f)$ is the designed sensitivity of either Virgo (V), LIGO (L),
Advanced LIGO (AL) or of the planned resonant detector Dual. Of course,
the SNR is inversely proportional to the distance from the source and the
values reported in Table~\ref{table:GWamplitude} indicate that, while an
instability developing in a rapidly rotating star in our Galaxy would
yield an extremely strong signal, the SNR is still expected to be ${\cal
O}(1)$ also for a source as far as about 10 Mpc if measured by Advanced
LIGO.  A different representation of the results summarised in
Table~\ref{table:GWamplitude} is offered in Fig.~\ref{fig:SNRLIGO}, which
shows a spectral comparison between the designed sensitivity
$h_{rms}(f)\equiv\sqrt{S_n(f)}$ of Virgo, LIGO and Advanced LIGO and the
power spectrum $|\tilde{h}_{+}(f)|f^{\frac{1}{2}}$ of the expected
signals for models U3 and U13.

It is interesting to note the significant differences in the power
spectrum of the two signals, with the one relative to model U3 having
a larger and narrower peak as a result of a more persistent bar. Yet,
the SNR for model U3 is significantly smaller than the one for both
models U11 and U13 and a number of different factors are behind this
somewhat surprising result. Firstly, while the bar is indeed more
persistent for U3, the amplitude of the bar distortion is larger for
models U11 and U13, thus larger is the corresponding
gravitational-wave signal. Secondly, the frequency of the
power-spectrum maximum is smaller for models U11 and U13 and thus
better fitting the sensitivities of the detectors.  Finally, the
idealised assumption that match-filtering techniques can be used at
all frequencies implies that all of the power spectrum contributes to
the final SNR and the large wings of the spectra of models U11 and U13
can significantly increase the SNR. Indeed, when evaluating
eq.~\eqref{eq:snr2} in a narrow window of 100 Hz around the peak, the
SNR of the different models is comparable.

Of course, a strong SNR is just a necessary condition for the detection
and the possibility of measuring the gravitational-wave signal from this
process will depend significantly on its event rate, which is still
largely uncertain. In the case in which the instability develops in a hot
protoneutron star resulting from the collapse of a stellar core, the
event rate is strictly related to the supernova event rate, which is of 1
or 2 supernova per century per galaxy. About 60\% of the remnants of
the explosion should be neutron stars, but the requirement of rapid
rotation in the progenitor makes the event rate of dynamical
instabilities considerably lower~\cite{KokkotasStergioulas2005}, although
such prospect may be more optimistic after recent studies~\cite{Ott:2006}. On the other hand, in
case the bar is produced as a result of a binary merger of neutron
stars, the most optimistic scenarios suggest that such mergers may occur
approximately once per year within a distance of about 50
Mpc~\cite{Kalogera:2003tn}. Finally, the event rate of the classical
scenario in which the instability is triggered in an old neutron star
spun up by accretion in a binary system, still remains difficult to
quantify.

\section{Conclusions}
\label{sec:conclusions}

We have presented accurate simulations of the dynamical bar-mode
instability in full General Relativity. An important motivation behind
this work is the need to go beyond the standard phenomenological
discussion of the instability and to find answers to important open
questions about its nonlinear dynamics. Among such open problems there are, for
instance, the determination of the role of the initial perturbation or of the
symmetry conditions, or the influence on the dynamics of the value of
the parameter $\beta$, or, most importantly, the determination of the
timescale of the persistence of the bar deformation once this is
fully developed. Clearly, this latter question is a very pressing one in
gravitational-wave astronomy, as it bears important consequences on the
detectability of the whole process.

In order to provide answers to these questions we have explored
the onset and development of the instability for a
large number of initial stellar models. These have been calculated as
stationary equilibrium solutions for axisymmetric and rapidly rotating
relativistic stars in polar coordinates. More specifically, the evolved models
represent relativistic polytropes with adiabatic index
$\Gamma=2$ and are members of a sequence having a constant amount of
differential rotation with $\hat{A}=1$ and a constant rest mass of
$M_0 \simeq 1.51\ M_{\odot}$.

The simulations have been carried out with the general-relativistic
hydrodynamics code {\tt Whisky}, in which the hydrodynamics equations
are written as finite differences on a Cartesian grid and solved using
HRSC schemes~\cite{Baiotti03a,Baiotti04a}. The Einstein, equations, on
the other hand, have been solved within the conformal traceless
formulation implemented within the \texttt{Cactus} computational
toolkit~\cite{Cactusweb}.

The main results of our analysis can be summarised as follows:
\textit{i)} An initial $m$=1 or $m$=2-mode perturbation can play a role
in determining the duration of the bar-mode deformation, but not the
growth time of the instability; the only exception to this is represented 
by models near the threshold; \textit{ii)} For moderately overcritical
models the use of a $\pi$-symmetry can radically change the dynamics and
extend considerably the persistence of the bar; this ceases to be true
for largely overcritical models; \textit{iii)} The persistence of the bar
is strongly dependent on the degree of overcriticality and is generically
of the order of the dynamical timescale; \textit{iv)} Generic nonlinear
mode-coupling effects (especially between the $m$=1 and the $m$=2 mode)
appear during the development of the instability and these can severely
limit the persistence of the bar deformation and eventually suppress the
{bar deformation}; \textit{v)} The dynamics of largely overcritical
models (\textit{i.e.} with $\beta \gg \beta_c$) are fully determined by
the excess of rotational energy and the bar deformation is very rapidly
suppressed through the conversion of kinetic energy into internal one.
Interestingly, a similar dynamics for the odd and even modes has been
observed also in the case of the low-$\beta$ instability in recent
Newtonian simulations~\cite{OuTohline2006}. In this case, however, the
growth rate of the $m$=1 mode is much smaller and hence the persistence
of the bar longer.

Finally, we have considered whether the classical Newtonian stability
analysis of Maclaurin spheroids for incompressible self-gravitating
fluids is accurate also for differentially rotating and relativistic
stars. Overall, we have found the perturbative predictions to be
surprisingly accurate in determining the threshold for the instability as
well as the complex eigenfrequencies for the unstable models.

While the features of the bar-mode instability discussed here are expected
to be rather generic, they have been deduced from the analysis of a small
region of the space of parameters.  Work is now in progress to extend the
present analysis by considering stellar models with different and larger
compactnesses and that are regulated by more realistic EOSs.

\begin{acknowledgments}

The results presented here have benefited from discussions with several
friends and colleagues. We are particularly grateful to Nils Andersson,
Harald Dimmelmeier, Ian Hawke, Kostas Kokkotas, Ewald M\"uller,
Alessandro Nagar, Christian Ott, Motoyuki Saijo, Bernard Schutz, David
Shoemaker, Masaru Shibata, Nikolaos Stergioulas, Joel Tohline and
Burkhard Zink. Support for this research comes also through the SFB-TR7
of the German DFG and through the OG51 of the Italian INFN.
All the computations were performed on the cluster for numerical
relativity \textit{``Albert''} at the University of Parma.

\end{acknowledgments}

\end{document}